%% file: acl_latex.tex
\documentclass[11pt]{article}

\usepackage{acl}
\usepackage{comment}
\usepackage{times}
\usepackage{latexsym}
\usepackage{amsmath} 
\usepackage{algorithm} 
\usepackage{algpseudocode}
\usepackage{amsmath}
\usepackage{xcolor}
\usepackage{tabularx}
\usepackage{booktabs}
\usepackage[T1]{fontenc}
\usepackage[utf8]{inputenc}
\usepackage{listings}
\usepackage{microtype}

\usepackage{graphicx}

\usepackage{hyperref}
\usepackage{url}
\usepackage{enumitem}
\usepackage{booktabs}
\usepackage{multirow}

\usepackage{graphicx}
\usepackage{subcaption}

\newcommand{\ToolName}{\textsc{HintPilot}}

\definecolor{mygreen}{rgb}{0,0.6,0}

\newcommand{\mytodoblue}[1]{\textcolor{blue}{\ding{46}~{\sf}~#1}}
\newcommand{\mytodored}[1]{\textcolor{red}{\ding{46}~{\sf}~#1}}

\newcommand{\mytodoorange}[1]{\textcolor{orange}{\ding{46}~{\sf}~#1}}

\newif\ifshowcomments
\showcommentstrue

\ifshowcomments
\newcommand{\jiang}[1]{\mytodored{[jiang: #1]}}
\newcommand{\lin}[1]{\mytodored{[lin: #1]}}
\newcommand{\yao}[1]{\mytodoblue{[yao: #1]}}
\newcommand{\wang}[1]{\mytodoorange{[wang: #1]}}
\else
\newcommand{\jiang}[1]{}
\newcommand{\lin}[1]{}
\newcommand{\yao}[1]{}
\newcommand{\wang}[1]{}
\fi

\renewcommand{\smallskip}{}
\usepackage{enumitem}
\title{\ToolName{}: LLM-based Compiler Hint Synthesis for Code Optimization}
\author{
Hanyun Jiang$^{1}$, Peisen Yao$^{1,*}$, Kaiyue Li$^{1}$, Tingting Lin$^{1}$, Chengpeng Wang$^{2}$, Kui Ren$^{1}$ \\
$^{1}$The State Key Laboratory of Blockchain and Data Security, Zhejiang University \\
$^{2}$Purdue University \\
{\footnotesize \texttt{\{jhanyun,pyaoaa,kaiyue,polariso,kuiren\}@zju.edu.cn}} \\
{\footnotesize \texttt{wang6590@purdue.edu}}
}

\begin{document}
\maketitle
\begin{abstract}
Code optimization remains a core objective in software development, yet modern compilers struggle to navigate the enormous optimization spaces. While recent research has looked into employing large language models (LLMs) to optimize source code directly, these techniques can introduce semantic errors and miss fine-grained compiler-level optimization opportunities. We present \ToolName{}, which bridges LLM-based reasoning with traditional compiler infrastructures via synthesizing \emph{compiler hints}—annotations that steer compiler behavior.
\ToolName{} employs retrieval-augmented synthesis over compiler documentation and applies profiling-guided iterative refinement to synthesize semantics-preserving and effective hints. Upon PolyBench and HumanEval-CPP benchmarks, \ToolName{} achieves up to 6.88$\times$ geometric mean speedup over \texttt{-Ofast} while preserving program correctness. Our code is available at \url{https://github.com/ZJU-PL/hintpilot}.
\end{abstract}

\input{1.intro}

\input{2.background}

\input{3.method}
\input{4.experi}
\input{6.related}

\input{7.conclu}

\bibstyle{acl_natbib}
\bibliography{ref,custom,logic,symexe,tmp}

\clearpage
\appendix
\input{appendix}

\end{document}

%% file: 1.intro.tex
\section{Introduction}

Code optimization is fundamental to software performance, directly affecting execution speed, energy efficiency, and resource utilization across domains ranging from embedded systems to large-scale cloud computing~\cite{8357388, Garg_2022}. Traditionally, compilers have served as the primary vehicle for optimization, relying on expert-crafted heuristics applied during the translation from high-level source code to machine instructions. As modern software systems grow increasingly complex and heterogeneous, however, selecting effective optimization strategies has become both labor-intensive and insufficiently adaptive, motivating a shift from static designs toward dynamic, data-driven solutions~\cite{scott}.

\begin{figure}[t]
    \centering
    \begin{subfigure}{\linewidth}
        \centering
        \includegraphics[width=\linewidth]{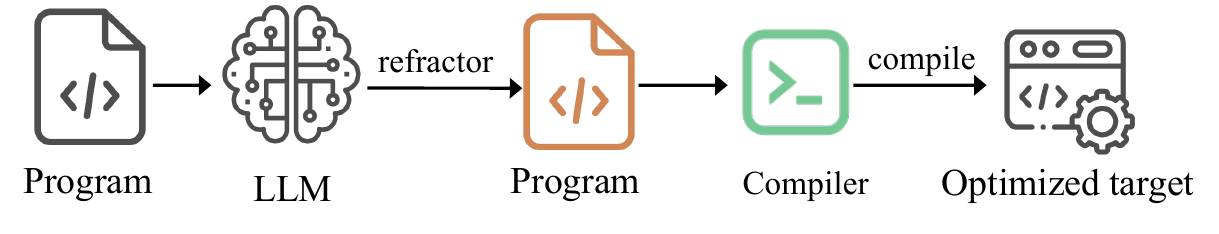}
      %  \vspace{-8mm}
        \caption{Paradigm I: Transforming code using LLMs}
     %   \vspace{4mm}
        \label{fig:cmp_direct}
    \end{subfigure}
    \begin{subfigure}{\linewidth}
        \centering
        \includegraphics[width=\linewidth]{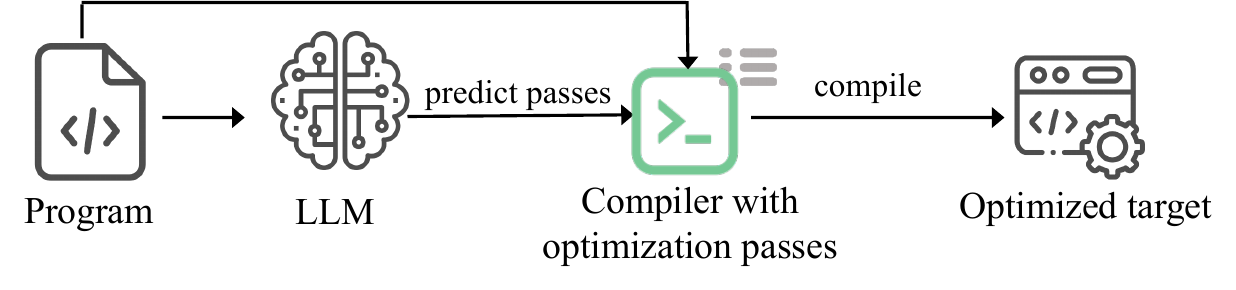}
      %  \vspace{-8mm}
        \caption{Paradigm II: Selecting compiler passes globally}
     %   \vspace{4mm}
        \label{fig:cmp_m2}
    \end{subfigure}
    \begin{subfigure}{\linewidth}    
        \centering
        \includegraphics[width=\linewidth]{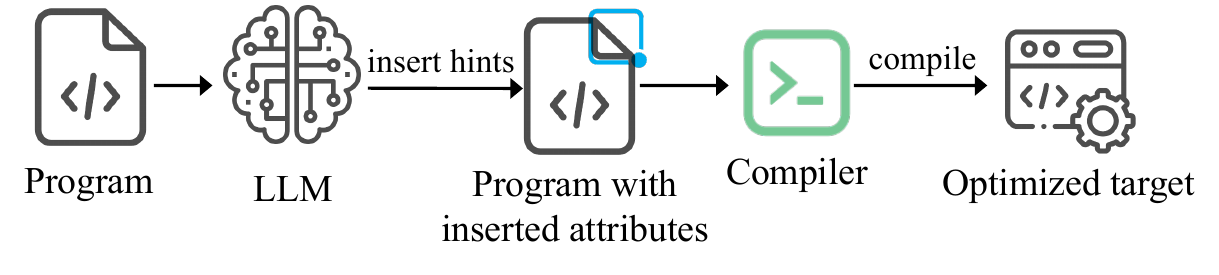}
      %  \vspace{-8mm}
        \caption{Our solution: Synthesizing compiler hints within code}
        \label{fig:cmp_m3}
    \end{subfigure}
    \caption{The comparison between different paradigms of LLM-based code optimization}
    \label{fig:comparison}
   % \vspace{-5mm}
\end{figure}

Recent advances in large language models (LLMs) have shown promise in generating performant code~\cite{Gong_2025, Dong_2025}. Nevertheless, ensuring correctness alongside performance remains a fundamental challenge~\cite{ecco_2024, Yang_2024}. Existing LLM-based optimization approaches often operate by directly modifying source code to improve efficiency~\cite{pie24, Gao_2024, Zhao_2025} (See Fig.~\ref{fig:comparison}(a)). While such methods can yield performance gains, they typically rely on invasive transformations that risk violating program semantics.
Alternative work explores LLM-guided selection of compiler passes or flags without modifying source code~\cite{llmcompiler} (See Fig.~\ref{fig:comparison}(b)). Still, they usually apply a single global optimization configuration to the entire program, missing fine-grained opportunities and overlooking the performance impact of functions, variables, and non-loop statements.

\begin{figure}[t]
    \centering
    \begin{subfigure}{\linewidth}
        \includegraphics[width=\linewidth]{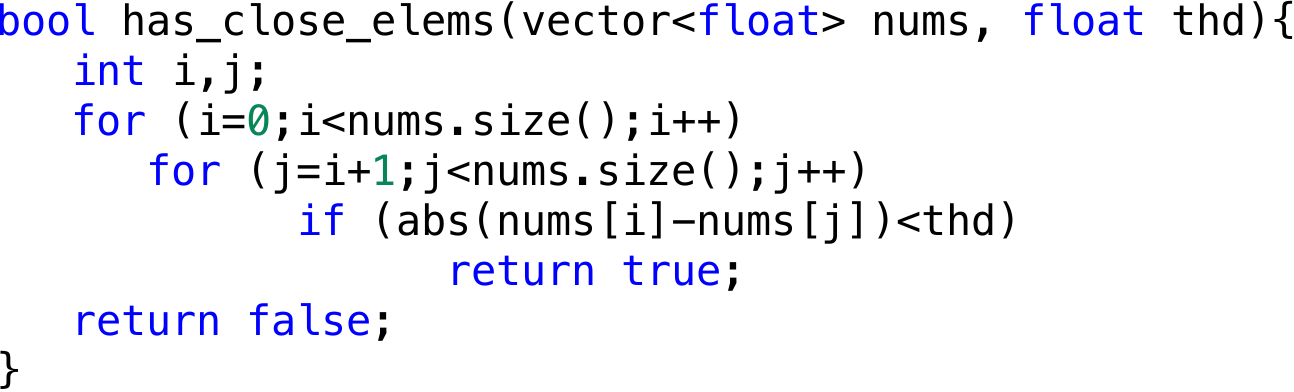}
       % \vspace{-8mm}
        \caption{The original code before optimization}
        \label{fig:exp_in}
    \end{subfigure}
    \begin{subfigure}{\linewidth}
        \includegraphics[width=\linewidth]{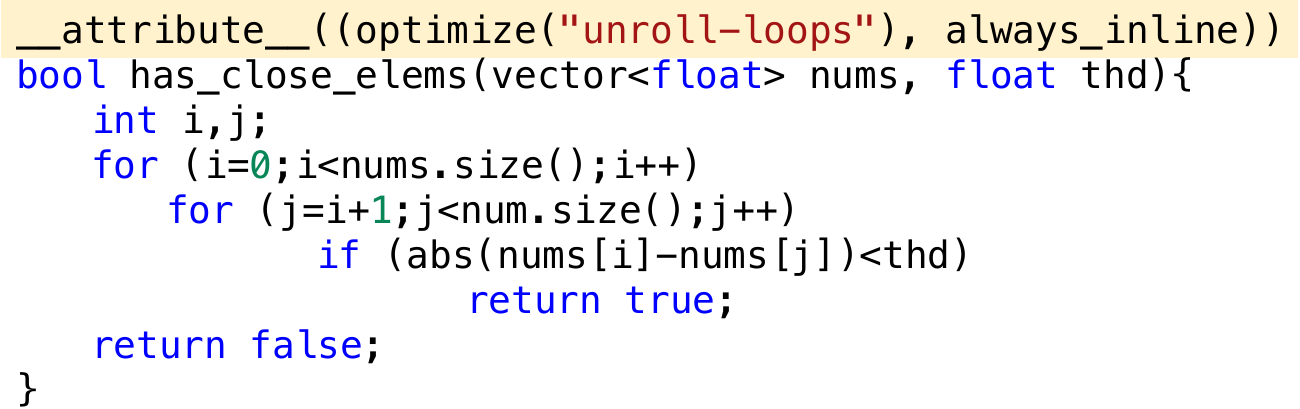}
       % \vspace{-3mm}
        \caption{The optimized code with the compiler hints}
    \label{fig:exp_out}
    \end{subfigure}
    \caption{An example of code optimization by synthesizing compiler hints. \texttt{optimize("unroll-loops")} unrolls the loops in the function. \texttt{always\_inline} inlines the function when it is called
    } 
    \label{fig:exp}
  %  \vspace{-3mm}
\end{figure}

To fill this gap, we introduce a new paradigm for code optimization by synthesizing compiler hints, which reconciles the flexibility of LLM-based code reasoning with the reliability guarantees of traditional compilers (see Fig.~\ref{fig:comparison}(c)). Specifically, compiler hints are annotations attached to functions, statements, or classes of a program, with common categories summarized in Table~\ref{tab:hint_categories}.
Notably, applying a small set of such hints to a program can yield an optimized version. For example, the program in Fig.~\ref{fig:exp}(b) with the compiler hint achieves a 1.47$\times$ speedup over the one in Fig.~\ref{fig:exp}(a), with correctness preserved by construction.
Leveraging such compiler hints, we can achieve code optimization with two key benefits. First, we optimize via declarative hints rather than direct code rewriting. Restricting the model’s output to compiler-validated annotations ensures functional correctness. Second, hint-based optimization enables fine-grained, location-specific control. Unlike global compiler flags, hints can be selectively applied to individual program elements, allowing fine-grained, targeted, and context-aware performance tuning.

\begin{table}[t]
\centering
\small
\caption{Examples of compiler hints}
\label{tab:hint_categories}
\setlength{\tabcolsep}{10pt}
\renewcommand{\arraystretch}{1.15}
\begin{tabularx}{\linewidth}{@{}lX@{}}
\toprule
\textbf{Category} & \textbf{Representatives} \\
\midrule
Storage control & \texttt{section} \\
Optimization control & \texttt{used}, \texttt{unused} \\
Memory layout & \texttt{packed} \\
Alignment control & \texttt{aligned(\textit{n})} \\
Warning handling & \texttt{warn\_unused\_result} \\
Visibility control & \texttt{visibility("hidden")} \\
Entry control & \texttt{naked}, \texttt{interrupt} \\
Calling convention & \texttt{cdecl}, \texttt{stdcall} \\
Constructor/Destructor & \texttt{constructor}, \texttt{destructor} \\
\bottomrule
\end{tabularx}
%\vspace{-3mm}
\end{table}

Despite these advantages, determining where and how to apply compiler hints remains challenging, as their effects depend on complex, non-local interactions among program characteristics and compiler behavior. 
First, synthesized hints should preserve program semantics. Incorrect usage can lead to undefined behavior or wrong results.  Second, hint synthesis should be precise to avoid compiler rejections and to produce hints that yield measurable performance improvements.

To address this, we formulate the compiler hint synthesis for the first time as a structured prediction problem,
where LLMs generate structured outputs based on external compiler documentation that contains the semantics and usage constraints of available hints.
We instantiate this formulation in \ToolName{}, a framework for LLM-based hint synthesis. 
To ensure semantic preservation, we curate a knowledge base of side-effect-free hints and apply preprocessing to filter out unsafe annotations. Building on this knowledge base, we integrate retrieval-augmented generation with execution-guided feedback to synthesize and iteratively refine contextually appropriate hints, thereby improving performance while preserving correctness. 

We evaluate \ToolName{} upon a diverse benchmark suite, including \text{PolyBench}~\cite{polybench_sourceforge} and \text{HumanEval\_CPP}~\cite{zheng2024humanevakx}, which cover both structured numerical kernels and general-purpose C++ programs. Our experiments show that \ToolName{} consistently outperforms standard compiler optimization baselines
under different prompting strategies. In particular, it achieves geometric mean speedups of up to 3.53$\times$ over \texttt{-O3} and 6.88$\times$ over \texttt{-Ofast}. We further compare against the LLM-Compiler-based optimization-pass selection baseline~\cite{llmcompiler} and show that our method achieves higher performance. In addition, ablation studies demonstrate that the observed performance gains are attributable to \ToolName{}'s ability to identify and exploit non-local optimization opportunities, such as interactions across functions or program regions, which are difficult for traditional heuristic-driven optimization strategies to capture.

%% file: 2.background.tex
\begin{figure*}[t]
    \centering
    \includegraphics[width=\textwidth]{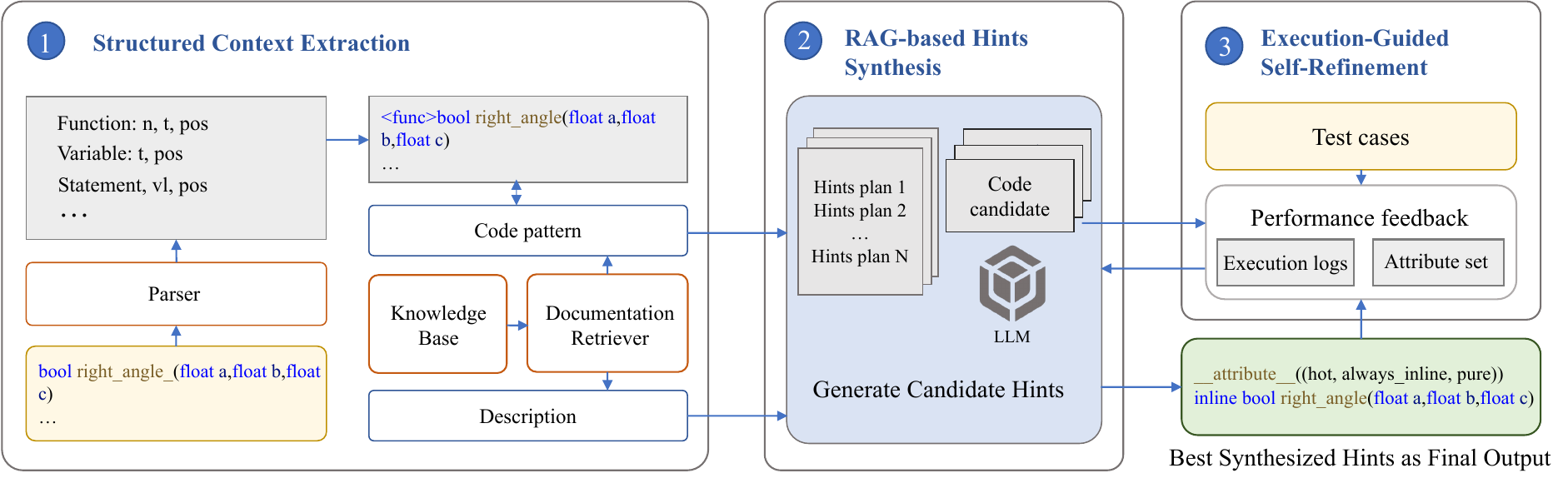}
    \caption{The workflow of \ToolName{}}
    \label{fig:overview}
  %  \vspace{-3mm}
\end{figure*}

\section{Preliminaries}

\noindent \textbf{Compiler Hints}.
Modern compilers, such as GCC, expose hints as lightweight annotations to convey developer intent to the compiler. By attaching hints to functions, loops, or variables, programmers can communicate semantic and performance-related properties that are difficult or costly for the compiler to infer reliably. Typical examples include performance-oriented hints (e.g., \texttt{hot}), branch-related cues (e.g., \texttt{likely}), and memory-behavior directives (e.g., \texttt{prefetch}).

Compiler hints occupy a distinct position in the optimization stack. Unlike source-level code transformations, hints do not alter control flow or data flow; instead, they declaratively constrain or guide the compiler’s internal optimization decisions. Compared to global optimization flags, hints can be selectively applied to individual program elements, enabling fine-grained, localized, and context-sensitive tuning. In this sense, hints define a ``safe knob surface'' for performance tuning.

\smallskip
\noindent \textbf{Compiler Hint Synthesis.}
Based on the observation of compiler hints,
we formulate the problem of compiler hint synthesis for code optimization.
Let $P$ be a program to be optimized and $\mathcal{I}$ be a set of inputs of $P$.
For an input $a \in \mathcal{I}$, let $t(P,a)$ denote the execution time of running
program $P$ on $a$ under a fixed evaluation setting.
Let $\mathcal{L}(P)$ denote the set of valid insertion locations in $P$, and let $\mathcal{H}$ denote the set of
available compiler hints that are intended to preserve program semantics.
A \emph{hint assignment} $S$ is modeled as a (partial) mapping $S : \mathcal{L}(P) \rightarrow \mathcal{H}$,
which assigns compiler hints to a selected subset of program locations.
We denote by $P \oplus S$ the program obtained by augmenting $P$ with the compiler
hints specified by $S$ at their corresponding locations.

Based on these concepts, we formulate the problem as follows: 
Given $P$ and $\mathcal{I}$, the goal is to find an assignment $S^{\ast}$ that minimizes the overall execution time over the input set $\mathcal{I}$:
\begin{align*}
S^\ast = \arg\min_{S} \ \sum_{a \in \mathcal{I}} t(P \oplus S, a).    
\end{align*}
However, effective compiler hint synthesis requires more than syntactic correctness; it demands semantic understanding of both program behavior and compiler internals. Fortunately, LLMs offer a promising foundation for this task, given their demonstrated strengths in code summarization~\cite{fang2024esale}, intent inference~\cite{wang2025boosting, ruan2024specrover}, and edit generation~\cite{dong2025survey}.
Nevertheless, direct prompting cannot solve the problem end-to-end. First, inserted hints must preserve program semantics. Incorrect annotations, for example, marking a side-effecting function as \texttt{const}, can induce undefined behavior or miscompilation. Second, compiler hints constitute long-tail knowledge in LLM training corpora. As a result, models may hallucinate hints, apply them incorrectly, or generate annotations that are syntactically valid but semantically vacuous. Such outputs may fail to influence optimization decisions, yielding no performance benefit.
To resolve the above challenges, we introduce a framework for compiler hint synthesis, named \ToolName{}, detailed in Sec.~\ref{methodology}.

%% file: 3.method.tex
\section{Our Solution: \ToolName}
\label{methodology}
Figure~\ref{fig:overview} depicts the workflow of \ToolName{}. Following existing studies~\cite{pie24}, it takes a target program and a set of test cases as input. Technically, \ToolName{} synthesizes compiler hints for code optimization through three phases.
First, it parses the program to identify valid insertion sites via \emph{structured context extraction}. Second, during the stage of \emph{RAG-based hint synthesis}, \ToolName{} retrieves relevant hint descriptions and usage examples from the knowledge base, which are incorporated into a prompt that guides the LLM to generate a sequence of candidate hints. Third, \ToolName{} applies the generated hints, compiles the program, and executes the target code.
If compilation or testing fails or performance degrades, \ToolName{} performs \emph{execution-guided self-refinement}, leveraging the diagnostics to guide iterative regeneration.

Notably, \ToolName{} should exclude compiler hints that may potentially alter program semantics. Hence, to ensure program correctness, we preprocess compiler documentation and construct a knowledge base of semantic-preserving compiler hints. In what follows, we first describe the construction of this knowledge base (Sec.~\ref{subsec:kb}) and then present the technical details of each stage of the framework (Sec.~\ref{subsec:extract}$\sim$Sec.~\ref{subsec:feedback}).

\subsection{Knowledge Base Construction}
\label{subsec:kb}
Before code optimization, we construct a structured knowledge base that maps compiler hints to their semantic intent and usage patterns. This knowledge base provides the semantic grounding that later components require to generate valid, context-sensitive hint insertions.

\smallskip
\noindent  \textbf{Selecting Semantics-Preserving Hints.}
We extract hint descriptions from the official documentation for compilers, such as GCC, which we use in the evaluation to specify each hint’s intended use, applicability conditions, and semantic implications. From this corpus, we conservatively select a curated subset of 46 hints that do not affect a program’s observable behavior. Hints that introduce side effects or alter semantics are excluded. The retained hints serve solely as declarative guidance to the compiler, enabling optimizations without compromising functional correctness.

%\wang{Notably, the knowledge base construction can be generalized to different compilers and their documentations. xxxxx. }\jiang{fixed}
%Notably, our knowledge base is driven by compiler documentation and a hint schema. By replacing the documentation source and updating the hint mapping, the same pipeline can be instantiated for other compilers like Clang/LLVM, ICC/ICX, MSVC, and across architectures like x86, ARM, RISC-V, since such hints are expressed at the source level and are broadly portable, even though specific directives may differ.
Notably, our knowledge base is structured around a general hint schema and is documentation-driven. This design enables straightforward adaptation to other compilers—such as Clang across architectures including x86, ARM, and RISC-V.
% Since these hints are typically expressed at the source level, the methodology remains applicable despite syntactic and directive-level differences across toolchains.

\smallskip
\noindent  \textbf{Extracting Optimization Patterns.} 
We construct an external knowledge base from official compiler documentation that maps abstract optimization patterns to concrete hint usages. For each hint, we extract its full description. When descriptions are overly long, we use Gemini-2.5 to produce concise summaries of the key information. We also carefully identify and extract the applicable use cases and collect official code examples when available. For hints without official usage programs, we prompt Gemini-2.5 to generate examples and manually verify their correctness. At inference time, we retrieve contextually relevant entries and provide them to the model as explicit semantic grounding, thereby reducing the risk of generating invalid or hallucinated hints.

\subsection{Structured Context Extraction}
\label{subsec:extract}
To localize the valid program location for compiler hints,
\ToolName{} performs structure-aware analysis of the input program to identify valid and promising locations for hint insertion. Rather than treating code as unstructured text, \ToolName{} parses the source into a structured abstraction that exposes functions, variables, and statements along with their locations and types.

The abstraction is obtained using the GCC parser, following the representation in~\cite{atlas25}:
\begin{align*}
    C = (F_{n,t,pos}, V_{t,pos}, S_{vl,pos}),
\end{align*}
where $F_{n,t,pos}$ denotes function metadata including function name $n$, return type $t$, and definition location $pos$; $V_{t,pos}$ denotes variable types and declaration locations; and $S_{vl,pos}$ denotes statements, the variables they reference, and their locations. 

This abstraction enables \ToolName{} to (i) identify syntactically valid insertion points such as function definitions and loop headers, and (ii) isolate the minimal structural context required for subsequent retrieval and generation. By constraining the search space to structurally valid regions, this component reduces noise and improves the reliability of downstream LLM guidance.
% \\wang{For example, for the program in Fig.~\ref{fig:exp}(a), xxx}\jiang{this step is used to generate input program with markers, so I elaborate in next section with examples}

\subsection{RAG-based Hint Synthesis}
\label{subsec:rag}

Building on the extracted structural context, the second component uses retrieval-augmented generation (RAG) to guide the synthesis of hints. \ToolName{} combines code structure analysis with semantic grounding from the knowledge base to produce context-aware optimization hints. We present an example of the retrieved content in Appendix~\ref{app:case}.

\begin{figure}[t]
    \centering
    \includegraphics[width=0.9\linewidth]{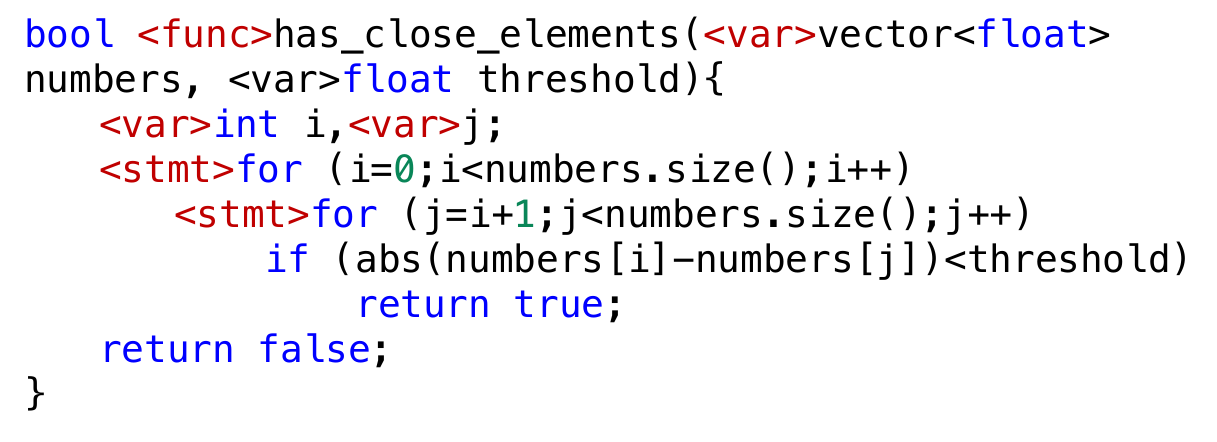}
    %\vspace{-1mm}
    \caption{Input program format}
    \label{fig:exp3}
   % \vspace{-5mm}
\end{figure}

\smallskip
\noindent  \textbf{Prompt Construction.}
The LLM is prompted with the structural abstraction $C$ and code to suggest possible hints. The prompt includes:

% analyze the target code and
\emph{Structural Features:} Observations on code structure based on $C$ highlighting their potential impact on program behavior with corresponding markers like \texttt{<var>}, \texttt{<stmt>}, and \texttt{<func>} at the positions given by $F_{n,t,pos}$, $V_{t,pos}$, and $S_{vl, pos}$. Figure~\ref{fig:exp3} demonstrates an example with the marker applied to source code, where possible insertion positions are marked for LLM.

\begin{figure}[t]
    \centering
    \includegraphics[width=\linewidth]{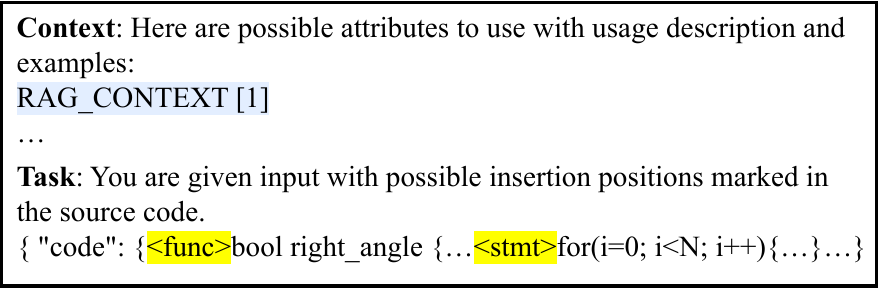}
    \caption{The prompt template for hint synthesis}
    \label{fig:prompt1}
  %  \vspace{-3mm}
\end{figure}

\emph{Optimization Recommendations.} A set of relevant hint information (\texttt{RAG\_CONTEXT} in Figure~\ref{fig:prompt1}) retrieved from the knowledge base, paired with concise applicability conditions to encourage correct usage in the given structural context. 
% \yao{too short}

\smallskip
\noindent  \textbf{Retrieval-Augmented Guidance.} To enhance LLM's contextual understanding, a RAG structure retrieves examples from a dataset, stored in a vector database. Each entry in the database contains:
\begin{itemize}[leftmargin=*, parsep=1pt, itemsep=1pt, topsep=2pt]
    \item Descriptions: Descriptions of the hints extracted from the official document, including their impact, usage, and conditions.
    \item Code pairs ($P_p$, $P_n$): Correct use examples of the hints and an example without the hints.
\end{itemize}

This retrieval mechanism addresses a common limitation of pretrained LLMs, which may lack precise compiler-specific knowledge and can misapply hints even when the intended optimization is reasonable. To further improve reliability, \ToolName{} adopts a parsing-and-planning workflow that first determines candidate insertion sites from $C$ and then synthesizes hints conditioned on retrieved evidence, instead of directly generating end-to-end hint-annotated code.

\subsection{Execution-Guided Self-Refinement}
\label{subsec:feedback}
\begin{figure}
    \centering
    \includegraphics[width=\linewidth]{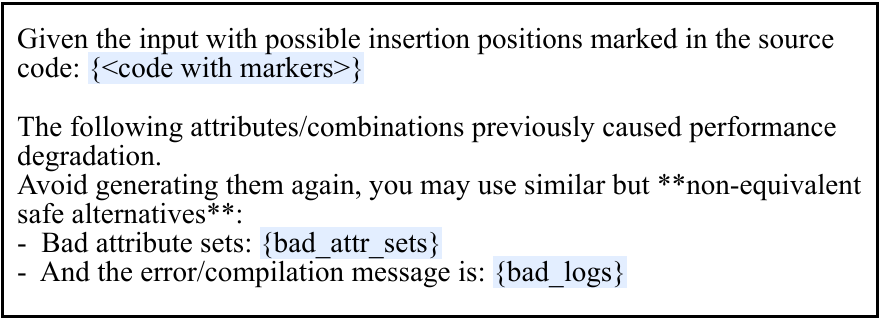}
    \caption{The prompt sketch for refinement}
    \label{fig:prompt2}
   % \vspace{-3mm}
\end{figure}

We incorporate additional test cases generated by LLMs following~\cite{pie24}. During the profiling phase, we utilize these cases to gather runtime information and identify potential performance bottlenecks. The model takes the program source code, profiling data, and compilation feedback as input. The iterative refinement process follows a three-step feedback loop: 
\begin{itemize}[leftmargin=*, parsep=1pt, itemsep=1pt, topsep=2pt]
    \item \textbf{Suggestion:} The model proposes five candidate sets of compiler hints for the target code, resulting in multiple feedback signals.
    \item \textbf{Execution:} The code is compiled with the proposed hints and benchmarked using test cases to obtain performance metrics.
    % LLM-generated 
    \item \textbf{Feedback:} The measured results, together with the applied hints, are fed back to the model to guide the next iteration. They are \texttt{bad hint sets} and \texttt{bad logs} in Figure~\ref{fig:prompt2}.
\end{itemize}

If the refined program passes all test cases and achieves a measurable speedup, it is evaluated on the official test suite. This feedback-guided loop enables \ToolName{} to adaptively explore the hint space, correcting invalid insertions and converging on performance-improving configurations.

%% file: 4.experi.tex
\section{Evaluation}
\label{sec:eval}
We evaluate \ToolName{} upon benchmarks to quantify its effectiveness.
This section details the experimental setup, evaluation results, and case studies. 

\subsection{Experimental Setup}
\noindent \textbf{Datasets.} We select two benchmarks to cover both numerical and general-purpose algorithms.
\text{Polybench}~\cite{polybench_sourceforge} comprises 34 numerical kernels essential to high-performance computing. It includes linear algebra operations (e.g., Cholesky decomposition), stencil computations (e.g., Jacobi), and dynamic programming, which is widely used in compiler optimization research.
\text{HumanEval\_CPP} is the C++ version of HumanEval-X ~\cite{zheng2024humanevakx} that extends HumanEval ~\cite{humaneval21} to multiple languages and includes 164
tasks. This dataset evaluates optimization across diverse algorithmic patterns, including sorting, searching, string manipulation, and graph algorithms.

\smallskip
\noindent  \textbf{Baselines.} 
We first compare \ToolName{} against two widely adopted compiler optimization flags. The first is \text{-O3}, the industry-standard optimization level that applies a comprehensive suite of optimizations to maximize performance. The second is \text{-Ofast}, a more aggressive optimization mode that includes all \text{-O3} optimizations along with additional transformations that may relax strict language standard compliance.
In addition, we compare \ToolName{} with LLM-Compiler~\cite{llmcompiler}, a state-of-the-art Meta-developed LLM for compiler optimization built on Code Llama. LLM-Compiler is trained on a range of compiler-centric tasks, such as compiler pass prediction and compiler emulation.
We use the LLM-Compiler-13B variant to predict optimization flags for code optimization.

\noindent \textbf{Models.}
We evaluate a diverse set of models, including both proprietary APIs and open-weight architectures.
%to assess generalization across reasoning-intensive and conventional code generation tasks. 
The evaluated models comprise Qwen3-Coder-Plus, GPT-5.2, Codestral-22B-v0.1, Qwen2.5-Coder-14B-Instruct, GPT-4o-mini, and Claude-Sonnet-4.5.
To ensure a fair comparison with the baseline llm-compiler, we configure \ToolName{} to use the same backbone model, CodeLlama-13B-Instruct, thereby isolating the impact of our method from model capacity. 

%\smallskip
\noindent  \textbf{Prompting Strategies}.
We investigate three prompting strategies for synthesizing compiler hints. In the zero-shot setting, the model is provided with source code and profiling information and instructed to generate compiler hints directly. To enforce the multi-step reasoning, we also adopt a chain-of-thought (CoT) prompting strategy following prior work~\cite{rapgen_2025}.
%, in which the model is guided to reason about performance characteristics before producing optimization hints. 
Finally, for few-shot prompting, we augment the prompt with a small number of in-context demonstrations~\cite{fewshot2020} that illustrate compiler hint insertion. Concretely, we include five representative examples from  the  knowledge base, curated using Gemini-3 as few-shot examples.%\jiang{added}

\noindent  \textbf{Metrics}. We use geometric mean speedup to measure code performance, defined as:
%\wang{$S_i$ looks unnecessary. What does $N$ mean?}
\begin{equation*}
\text{Speedup}_{\text{geo}} =  \sqrt[N]{\prod_{i=1}^{N} \frac{T_{\text{baseline}, i}}{T_{\text{method}, i}}} .
\end{equation*}
where $N$ means the total number of test cases used for a program. 

All experiments are conducted on a 32-core AMD EPYC 7543 server with 512 GB of RAM, running Ubuntu 22.04 and using GCC 13.3.0. Each experiment is repeated ten times to reduce measurement variance, and runtime results are averaged across independent runs.

\subsection{Main Results}
We evaluate the effectiveness of \ToolName{}
%on the PolyBench and HumanEval\_CPP benchmarks 
by comparing it against two representative baselines.
First, we compare with conventional compiler optimization levels, including \text{-O3} and \text{-Ofast}.
Second, we compare with \text{llm-compiler}, a state-of-the-art LLM for compiler optimization.

\begin{figure}[t]
    \centering
    \begin{subfigure}{0.49\linewidth}
        \includegraphics[width=\textwidth]{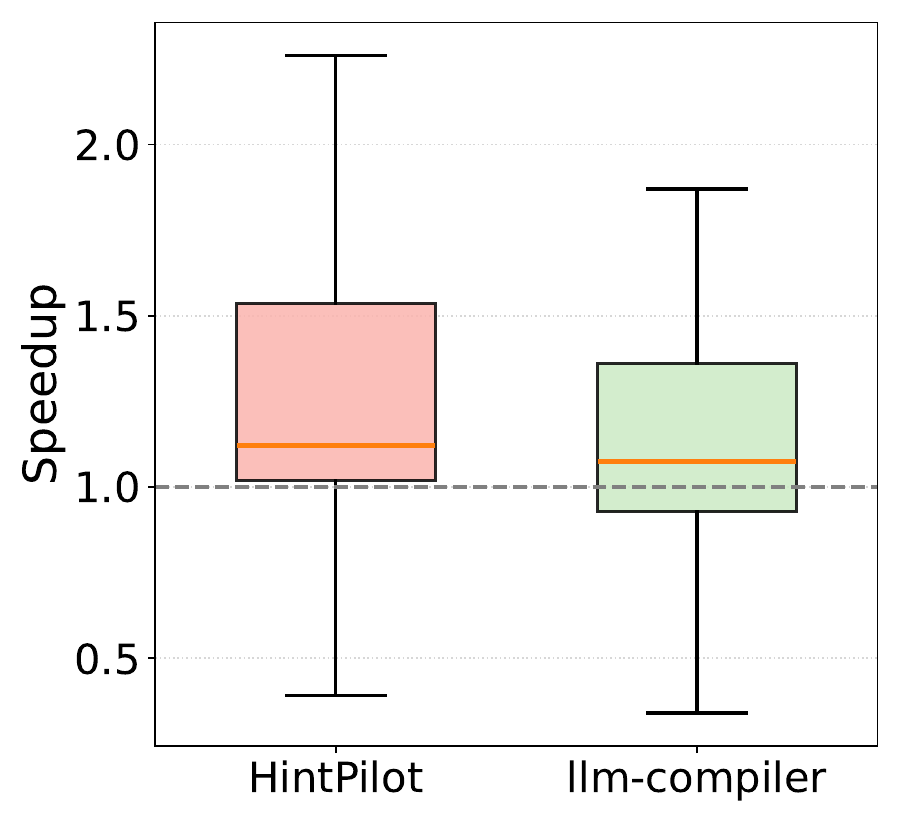}
        \caption{HumanEval\_CPP}
    \end{subfigure}
    \begin{subfigure}{0.49\linewidth}
        \includegraphics[width=\textwidth]{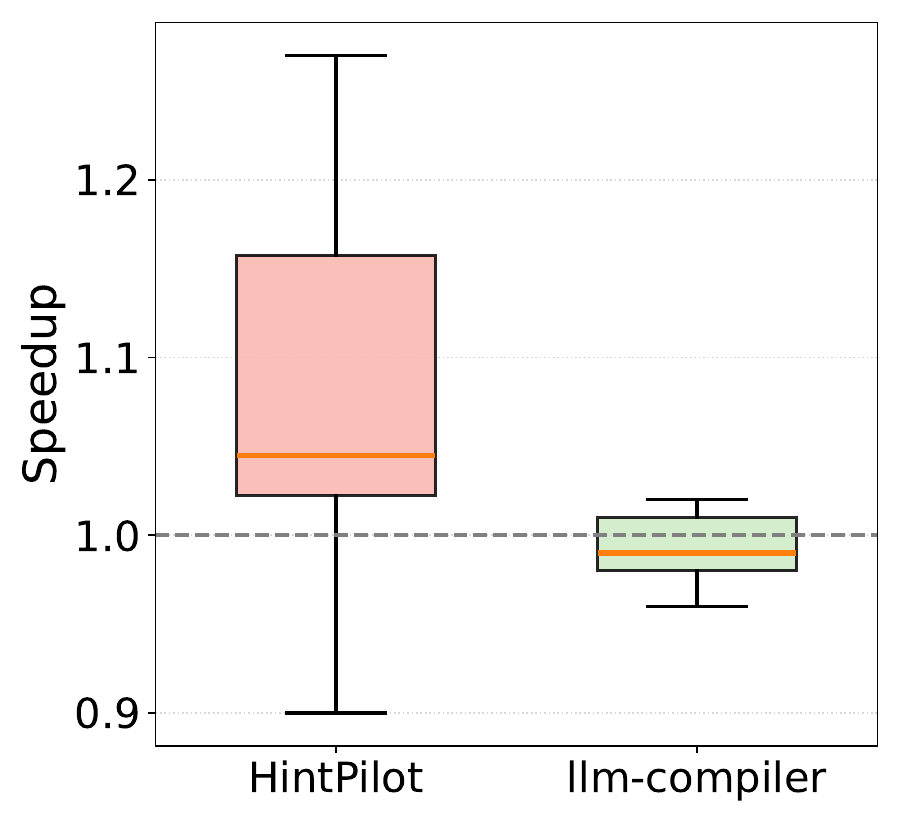}
        \caption{PolyBench}
    \end{subfigure}
    \caption{Boxplot of geometric mean speedup relative to O3 of \ToolName{}(Codallama-13b-instruct) compared to the llm-compiler-13b baseline.}
    \label{fig:llmcompiler}
    \vspace{-5mm}
\end{figure}

\noindent \textbf{Overall Effectiveness}. 
Figure~\ref{fig:llmcompiler} compares \ToolName{} with llm-compiler. As shown in the figure, \ToolName{} delivers higher speedups on both datasets, with its distribution consistently shifted upward relative to llm-compiler-13b. And the distribution is largely above 1, relative to the llm-compiler-13b, which may slow down the program.
These results suggest that generating compiler hints is a more effective optimization approach than directly predicting compiler passes for an entire program using LLMs.

\noindent \textbf{Comparison with Compiler Optimization Levels}.
As shown in Table \ref{tab:main_results}, \ToolName{} consistently outperforms the standard compiler optimization levels across different backend LLMs. When equipped with the model \text{Qwen3-Coder-Plus}, our framework achieves remarkable geometric mean speedups of 3.53$\times$ on \text{HumanEval\_CPP} and 2.10$\times$ on \text{Polybench} relative to \text{-O3}. Even compared to the aggressive \text{-Ofast} optimization level, which relaxes strict compliance with the standard for speed, \ToolName{} still delivers substantial gains, achieving 6.88$\times$ and 1.63$\times$ speedups, respectively. This demonstrates that our method effectively identifies fine-grained optimization opportunities that traditional compiler heuristics miss.

\noindent \textbf{Comparison with llm-compiler-13b}.
%\yao{speed here} 
We further present comparison results for \ToolName{} (using the model CodeLlama-13B-Instruct) against llm-compiler-13b (for predicting compiler flags).
Figure~\ref{fig:llmcompiler} reports the distribution of speedup rates and the proportion of programs that exhibit actual performance gains. \ToolName{} consistently outperforms \text{llm-compiler-13b}, achieving both higher average speedups and greater consistency across benchmarks. These results indicate that our method yields not only larger but also more reliable performance improvements.

\begin{table*}[t]
\caption{Geometric mean speedup compared with O3 and Ofast. $T$ refers to the maximum number of iterations. Here, $T$ denotes the maximum number of iterations and $N$ denotes the number of candidates. We first report results for different models with $T=2$ and $N=5$, and then analyze the impact of varying $T$ and $N$}
\label{tab:main_results}
\centering
\small
\begin{tabular}{lcccc}
\toprule

\multirow{2}{*}{\textbf{Optimization Option}} & \multicolumn{2}{c}{\textbf{O3}} & \multicolumn{2}{c}{\textbf{Ofast}} \\
\cmidrule(lr){2-3} \cmidrule(lr){4-5} 

& \textbf{HumanEval\_CPP} & \textbf{PolyBench} & \textbf{HumanEval\_CPP} & \textbf{PolyBench} \\
\midrule    

Qwen3-Coder-Plus & 3.53$\times$ & 2.10$\times$ & 6.88$\times$ & 1.63$\times$ \\
GPT-5.2 & 1.41$\times$ & 1.51$\times$ & 1.55$\times$ & 1.49$\times$ \\
Codestral-22B-v0.1 & 1.08$\times$ & 1.21$\times$ & 1.23$\times$ & 2.17$\times$ \\
Qwen2.5-Coder-14B-Instruct & 2.04$\times$ & 1.10$\times$ & 1.84$\times$ & 1.25$\times$ \\
GPT-4o-mini & 1.27$\times$ & 1.26$\times$ & 1.86$\times$ & 1.23$\times$ \\
Claude-Sonnet-4.5 & 2.88$\times$ & 1.34$\times$ & 3.87$\times$ & 1.35$\times$ \\
\midrule

Qwen3-Coder-Plus($T=2$, $N=1$) & 1.20$\times$ & 1.06$\times$ & 1.16$\times$ & 1.20$\times$ \\
Qwen3-Coder-Plus($T=3$, $N=1$) & 1.18$\times$ & 1.04$\times$ & 1.17$\times$ & 1.06$\times$ \\
Qwen3-Coder-Plus($T=2$, $N=3$) & 1.30$\times$ & 1.30$\times$ & 1.41$\times$ & 1.16$\times$ \\

\bottomrule
\end{tabular}
%\vspace{-3mm}
\end{table*}

\smallskip
\subsection{Ablation Study}
We conducted an ablation study comparing Zero-shot, Chain-of-Thought (CoT), and CoT + Few-shot strategies, as summarized in Table~\ref{tab:ablation}. The results reveal a clear progressive improvement.

\smallskip
\noindent  \textbf{Impact of Prompting Strategies}. Zero-shot yields only modest gains ($1.34\times$ on Humaneval\_CPP and $1.12\times$ on PolyBench), likely due to limited compiler-specific knowledge. Adding CoT improves Humaneval\_CPP by $1.62\times$, suggesting that reasoning helps identify optimization-friendly structures, such as dependency-free loops. CoT + Few-shot further boosts Humaneval\_CPP to $2.10\times$, but does not help PolyBench ($1.15\times$ vs. $1.17\times$ with CoT), implying that fixed examples can introduce bias or noise across diverse numerical kernels.

In contrast, \ToolName{} achieves a clear leap to $3.53\times$ on Humaneval\_CPP and $2.10\times$ on PolyBench. Moreover, CoT alone reduces the PolyBench compilation rate to $73.52\%$, whereas \ToolName{} recovers it to $80.00\%$ and achieves the highest speedup with few regressions. Overall, these results support the conclusion that our solution retrieves precise, context-aware patterns beyond those achievable with a static few-shot prompting strategy.

\smallskip
\noindent \textbf{Impact of Selected Models}. 
As shown in Table~\ref{tab:main_results},
we also observe that the choice of backbone LLM significantly affects optimization quality. The \text{Qwen3-Coder-Plus} and \text{Claude-Sonnet-4.5} models generally outperform smaller models such as \text{Codestral-22B}, validating that stronger reasoning capabilities in the base model translate into more effective compiler hints. %\yao{non concrete data?}

\smallskip
\noindent \textbf{Selection of Iteration and Candidate Numbers}. 
The ablation results in Table~\ref{tab:main_results} underscore the effectiveness of our execution-guided self-refinement. Increasing the candidate pool size $N$ from 1 to 3 (with $T=2$) yields a clear performance gain, suggesting that broader exploration of combinations of hints for richer execution feedback is key to discovering stronger configurations. However, increasing the number of refinement iterations leads to a slight performance drop, likely due to noise accumulation in the feedback signals.

\begin{table*}[t]
\centering
\small 
\caption{Analysis of compilation rate and efficiency on PolyBench and HumanEval\_CPP benchmarks. We use Qwen3-Coder-Plus. We report compilation rate (Comp.), speedup rate (Spd.), and geometric mean speedup relative to -O3. Best results among LLM-based methods are highlighted in bold.}
\label{tab:ablation}
\begin{tabular}{lcccccc}
\toprule
\multirow{2}{*}{\textbf{Method}} & \multicolumn{3}{c}{\textbf{HumanEval\_CPP}} & \multicolumn{3}{c}{\textbf{PolyBench}} \\
\cmidrule(lr){2-4} \cmidrule(lr){5-7}
 & \textbf{Comp. (\%)} & \textbf{Spd. (\%)} & \textbf{Speedup} & \textbf{Comp. (\%)} & \textbf{Spd. (\%)} & \textbf{Speedup} \\
\midrule

\quad Zero-shot & 94.51\% & 87.19\% & 1.34$\times$ & 78.04\% & 78.04\% & $1.12\times$ \\
\quad CoT & 97.56\% & 96.34\% & 1.62$\times$ & 73.52\% & 61.76\% & 1.17$\times$ \\
\quad CoT + Few-shot & 97.56\% & 96.95\% & $2.10\times$ & 70.58\% & 61.76\% & $1.15\times$ \\
\midrule

\quad \textbf{\ToolName{}} & 97.56\% & 97.56\% & \textbf{3.53$\times$} & 80.00\% & 80.00\% & \textbf{2.10$\times$} \\
\bottomrule
\end{tabular}
\end{table*}

\begin{figure*}[t]
    \centering
    \begin{subfigure}{0.24\linewidth}
        \includegraphics[width=\linewidth]{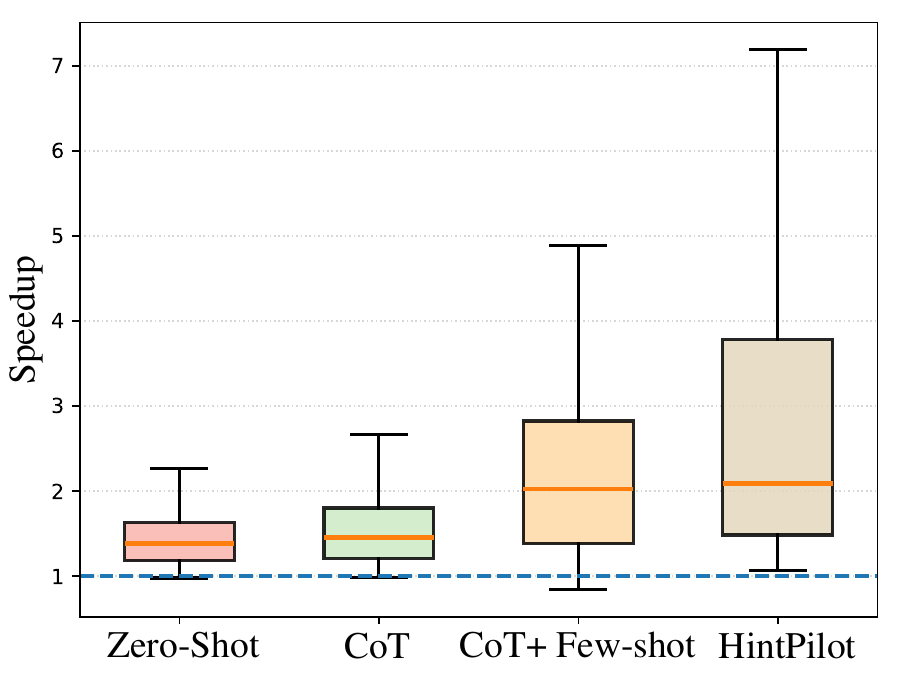}
        \caption{HumanEval\_CPP(O3)}
    \end{subfigure}
    \begin{subfigure}{0.24\linewidth}
         \includegraphics[width=\linewidth]{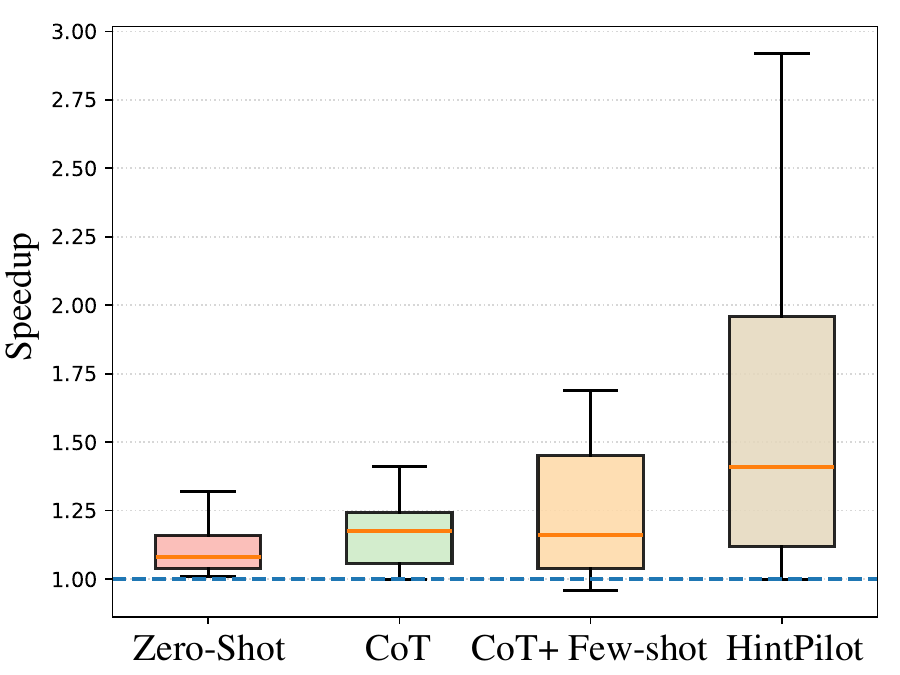}
        \caption{Polybench(O3)}
    \end{subfigure}
    \begin{subfigure}{0.24\linewidth}
         \includegraphics[width=\linewidth]{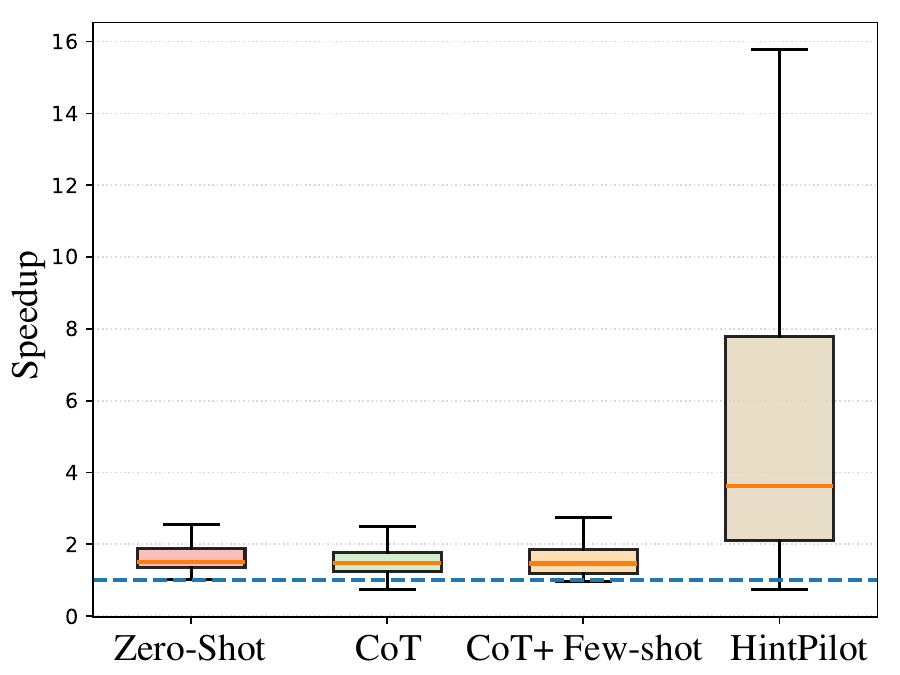}
        \caption{HumanEval\_CPP(Ofast)}
    \end{subfigure}
    \begin{subfigure}{0.24\linewidth}
         \includegraphics[width=\linewidth]{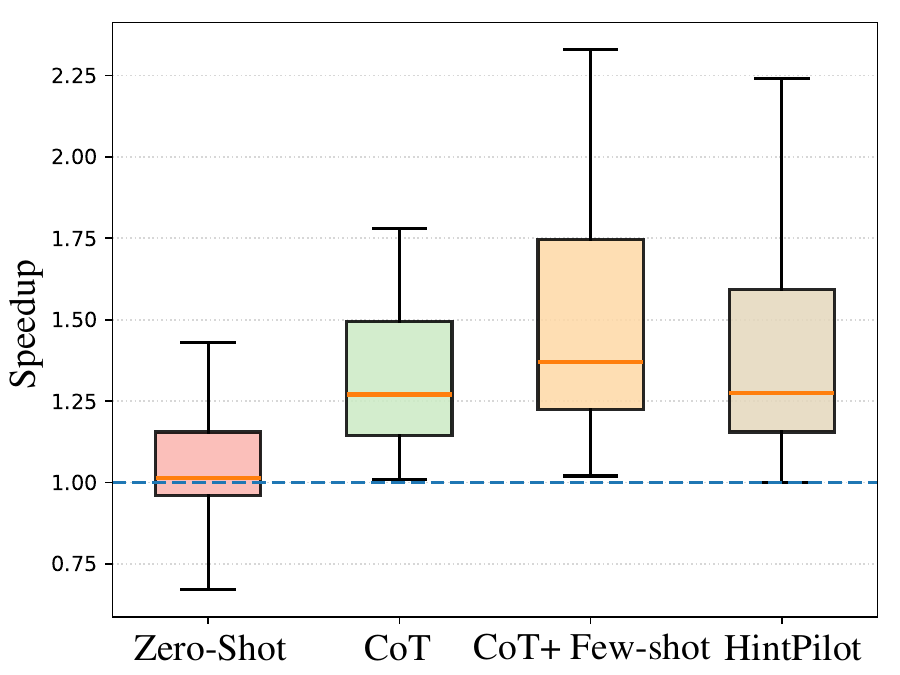}
        \caption{Polybench(Ofast)}
    \end{subfigure}
    
    \caption{Speedup boxplot of different methods across datasets and baselines}
    \label{fig:ablation-boxplot}
   % \vspace{-3mm}
\end{figure*}

\subsection{Failure Case Analysis}
\label{sec:fail}
%\wang{You could refer to \href{https://aclanthology.org/2024.findings-emnlp.217.pdf}{Link}}

Though the hints in our knowledge base are designed to be semantics-preserving when applied correctly, LLMs may still apply them incorrectly, such as inserting a hint at an inappropriate location or using an invalid syntax, which can lead to compilation errors. We therefore manually inspected cases where \ToolName{} failed to improve performance or caused compilation failures, and summarized the primary failure modes as follows:
\begin{itemize}[leftmargin=*, parsep=1pt, itemsep=1pt, topsep=2pt]
    \item \textbf{Syntax Hallucinations}: Despite retrieval augmentation, the model occasionally generates directives with invalid formats or hints unsupported by the specific compiler version. 
    \item \textbf{Contextual Mismatch}: Instances where hints are applied to incompatible scopes. For example, loop pragmas on non-loop statements or require compiler flags that were not active. 
    \item \textbf{Profiling Instability}: In rare cases, the feedback loop overfits to measurement noise, selecting candidates that offer negligible or unstable gains, as shown in Appendix ~\ref{case3}.
\end{itemize} 
Currently, 20\% of failures are due to syntax hallucinations, 79\% to context mismatch, and in rare cases (<1\%) to profiling instability. We attempted to mitigate syntax hallucinations using an LLM-based checker and stricter rule-based plan constraints; however, this tends to bias the model toward “safe” insertions (i.e., inserting in nearly all plausible positions) rather than producing the more selective hint insertions that drive performance gains. We therefore retain the current approach, since our primary goal is performance improvement, and the overall error rate remains low. Moreover, most syntax-related failures are cheaply detected at compile time, avoiding the need for high-cost test-case-based validation.
These observations highlight the need for future improvements in static hint verification and more robust profiling protocols to mitigate system noise.

%% file: 6.related.tex
\section{Related Work}
\label{sec:related}

\noindent  \textbf{LLM-based Code Optimization}.
Recent work on LLM-based code optimization falls into two main categories.
The first category comprises generative code refactoring approaches~\cite{Zhao_2025, Acharya2025, Gao_2024, supersonic, deepperf2022, rapgen_2025, codeoptimise}, which utilize LLMs to rewrite program structures. While benchmarks like PIE~\cite{pie24}, ECCO~\cite{ecco_2024}, Mercury~\cite{mercury2024}, EffiBench~\cite{Huangeffibench24}, and HumanEval~\cite{humaneval21} have demonstrated the potential of these models, they are primarily in Python, and such invasive changes often risk semantic drift. More recently, learning-based alignment strategies, such as EffiCoder~\cite{Huang2025efficoder}, PerfCodeGen~\cite{perfcodegen2024}, and ACECode~\cite{acecoder2024}, employ fine-tuning or reinforcement learning to align models with efficiency metrics, but incur high training costs.

The second category consists of compiler-centric techniques~\cite{llmcompiler, Merouani_2025, Lamour_2025, Baghdadi}, which integrate LLMs with internal compiler representations or cost models to guide transformations such as pass selection. While these methods offer a principled interface to the compiler, they typically lack fine-grained control at the source level.
In contrast, we introduce a lightweight, holistic paradigm spanning multiple granularities. Unlike heavy-weight, training-based, or invasive rewriting methods, we combine RAG with execution feedback to achieve significant performance gains while maintaining semantic correctness.

%\smallskip
\noindent \textbf{Machine Learning for Compilers}.
Compiler optimization involves navigating a high-dimensional transformation space to improve program performance. Traditionally, this process has relied on hand-crafted heuristics of compiler engineers. Recent work has explored the use of machine learning (ML) to automate heuristic design.
% enabling compilers to learn optimization strategies from data.
ML-based approaches have been applied to a range of compiler tasks, including vectorization~\cite{mendis2019ImitationLearning}, loop transformations such as unrolling and distribution~\cite{Stephenson2005, shalini-rl-loop-distribution-2022}, function inlining~\cite{trofin20MLGO}, and register allocation~\cite{das2020}. These methods typically train predictive models offline to replace fixed heuristics or to guide search-based optimization. Reinforcement learning has also been used to dynamically explore optimization sequences. Surveys such as~\cite{allamanis2018survey, wangSurvey2018} provide a broad overview of these techniques.
Prior work has primarily focused on selecting global compiler flags or tuning specific phases, such as register allocation. In contrast, our approach enables fine-grained synthesis of optimization hints, tailored to individual program components and capable of influencing multiple compiler phases.

%% file: 7.conclu.tex
\section{Conclusion}
\label{sec:conclu}
Software systems are increasingly complex and performance-critical, yet achieving effective optimization remains challenging. This work identifies an exciting, principled role for LLMs in addressing this challenge by synthesizing compiler hints—an interpretable and constrained interface between developers and compilers. We introduce \ToolName{}, a system that leverages LLMs to generate such hints, enabling compilers to uncover and exploit optimization opportunities without compromising correctness. Our results suggest that LLM-guided hint synthesis is a promising direction for improving code performance and making advanced compiler optimizations more accessible.
%to developers.

\section{Limitations}

While \ToolName{} demonstrates strong empirical performance across diverse benchmarks, several limitations remain.

\smallskip
\noindent  {\textbf{Scope of Optimization.}}
\ToolName{} focuses on source-code–level compiler hints that can be attached to localized program elements, such as functions, variables, and statements. As a result, it primarily targets fine-grained, local optimization opportunities exposed through compiler hints. More global optimization decisions, such as whole-program memory layout, interprocedural register allocation, or cross-module code placement, are outside the scope of the current framework because they are not directly controllable via localized hints. Extending \ToolName{} to reason about such global optimizations would require richer interfaces to the compiler and new forms of feedback beyond per-input runtime profiling.

\smallskip
\noindent  {\textbf{Reliance on Underlying LLMs.}}
The effectiveness of \ToolName{} depends on the reasoning and generalization capabilities of the underlying language model. Stronger models consistently produce higher-quality hint plans, particularly for non-local or uncommon optimization patterns, while smaller models are more prone to invalid or ineffective suggestions. Although retrieval-augmented grounding mitigates hallucinations and syntactic errors, it cannot fully compensate for the model's limited capacity. As LLMs continue to evolve, we expect \ToolName{} to benefit directly from improvements in model reasoning, code understanding, and long-context handling.

\smallskip
\noindent {\textbf{Benchmark Coverage and Generalization.}}
Our evaluation spans numerical kernels, algorithmic programming tasks, and competitive programming benchmarks, providing a broad view of \ToolName{} 's effectiveness. Nevertheless, these datasets do not fully capture all real-world optimization scenarios, such as large-scale industrial codebases, highly concurrent systems, or performance-critical I/O-intensive applications. Moreover, the input sets used for profiling and evaluation may not reflect the full diversity of production workloads. Expanding evaluation to additional datasets, architectures, and workload distributions is necessary to further assess robustness and generalization.

%% file: appendix.tex
\section{Algorithm of \ToolName{}}
\label{sec:appendix}
The algorithm of our approach is illustrated in Algorithm~\ref{alg:aptgen}, which is an iterative, retrieval-augmented code optimization framework driven by large language models (LLMs). Given an input program $P_{src}$, a domain-specific knowledge base $\mathcal{K}$, and a test suite $\mathcal{T}_{LLM}$, \ToolName{} searches for a semantically equivalent but higher-performing variant by automatically inserting optimization attributes (or hints) into the source code.

\smallskip 
\noindent \textbf{Initialization.}
The algorithm begins by profiling the original program $P_{src}$ on $\mathcal{T}_{LLM}$ to establish a baseline performance metric $M{best}$. The best-known program $P_{best}$ is initialized to $P_{src}$. In addition, \ToolName{} constructs a structural representation $S_{struct}$ of the source code using a compiler-based parser, extracting salient program elements such as functions, loops, and variables. An initially empty feedback history $H_{feedback}$ is maintained to accumulate information from prior optimization attempts.

\smallskip 
\noindent \textbf{Retrieval-Augmented Prompting.}
In each iteration, \ToolName{} first performs retrieval-augmented prompting. Using the structural summary $S_{struct}$ as a query, the retriever selects a set of relevant documents $D_{rag}$ from the knowledge base $\mathcal{K}$, including attribute specifications and usage examples. These retrieved artifacts, together with the source code and aggregated feedback from previous iterations, are used to construct a prompt $Prompt_t$ that contextualizes the optimization task for the LLM.

\begin{algorithm}[t]
\caption{\ToolName{}}
\label{alg:aptgen}
\small
\begin{algorithmic}
\Require Source Code $P_{src}$, Knowledge Base $\mathcal{K}$, Test Suite $\mathcal{T}_{LLM}$, Max Iterations $T$, Candidate Size $N$
\Ensure Optimized Code $P_{best}$

\State \textbf{Initialize:} $P_{best} \leftarrow P_{src}$, $M_{best} \leftarrow \textsc{Profile}(P_{src}, \mathcal{T}_{LLM})$
\State $H_{feedback} \leftarrow \emptyset$ \Comment{Initialize interaction history}
\State $S_{struct} \leftarrow \textsc{GCCParser}(P_{src})$ \Comment{Extract functions, loops, variables}

\For{$t = 1$ to $T$}
    \State \textcolor{blue}{\texttt{// Phase 1: Retrieval-Augmented Prompting}}
    \State $D_{rag} \leftarrow \textsc{Retrieve}(S_{struct}, \mathcal{K})$ \Comment{Fetch relevant attribute docs \& examples}
    \State $Prompt_t \leftarrow \textsc{ConstructPrompt}(P_{src}, S_{struct}, D_{rag}, H_{feedback})$
    
    \State \textcolor{blue}{\texttt{// Phase 2: Batch Generation of Plans}}
    \State $\mathcal{S}_{plans} \leftarrow \textsc{LLM}_{\pi}(Prompt_t, \text{samples}=N)$ \Comment{Generate $N$ independent insertion plans}
    
    \State $\mathcal{R}_{batch} \leftarrow \emptyset$
    \For{each plan $s_k \in \mathcal{S}_{plans}$}
        \State $P'_k \leftarrow \textsc{InsertHints}(P_{src}, s_k)$ \Comment{Deterministically apply attributes}
        \State $status_k, metric_k \leftarrow \textsc{Profile}(P'_k, \mathcal{T}_{LLM})$ \Comment{Compile and benchmark}
        \State $\mathcal{R}_{batch}.\text{add}(\{s_k, status_k, metric_k\})$
        
        \If{$status_k == \text{PASS} \land metric_k > M_{best}$}
            \State $P_{best} \leftarrow P'_k$
            \State $M_{best} \leftarrow metric_k$
        \EndIf
    \EndFor
    
    \State \textcolor{blue}{\texttt{// Phase 3: Feedback Aggregation}}
    \State $H_{feedback} \leftarrow \textsc{UpdateFeedback}(\mathcal{R}_{batch})$ \Comment{Summarize errors and perf gains}
\EndFor

\State \Return $P_{best}$
\end{algorithmic}
\end{algorithm}

\smallskip 
\noindent \textbf{Batch Plan Generation and Evaluation.}
Given $Prompt_t$, the LLM generates a batch of $N$ candidate insertion plans, each specifying a structured set of attribute insertions. Rather than directly emitting modified code, \ToolName{} deterministically applies each plan to the original program via \textsc{InsertHints}, yielding a candidate program $P'k$. Each candidate is then compiled and executed against $\mathcal{T}_{LLM}$ to assess both correctness and performance. Candidates that pass all tests and improve upon the current best metric are used to update $P_{best}$ and $M_{best}$.

\smallskip 
\noindent \textbf{Feedback Aggregation.}
After evaluating all candidates in the batch, \ToolName{} summarizes the observed outcomes—including compilation failures, runtime errors, and performance improvements—into an updated feedback history $H_{feedback}$. This feedback is fed into subsequent iterations, enabling the LLM to avoid previously unsuccessful patterns and refine future insertion plans.

\smallskip 
\noindent \textbf{Termination.}
The algorithm repeats this three-phase process for a fixed number of iterations $T$ and finally returns the best-performing program variant $P_{best}$ discovered during the search.

%\textbf{Implementation details} We implement AptGen using LangChain~\cite{} and vLLM~\cite{}. To mitigate potential hallucinations and ensure syntactic correctness, we adopt a two-stage generation strategy. Rather than generating annotated code directly, the model first produces a structured insertion plan, which is subsequently deterministically applied to the source code. We foster diversity in candidate generation by setting sampling parameters to temperature=1.0 and top-p=1.0. To guarantee reliable parsing of these plans, we enforce a strict JSON schema using the structured output features of the underlying frameworks. For the RAG component, we configure the retriever to fetch the top $k=4$ relevant documents as context.

%\section{Studied LLMs}

\section{Implementation Details}
%Certain compiler hints are only effective when paired with specific compiler flags. Accordingly, throughout our experiments, we keep the compiler options identical between the original program and the optimized variant to ensure a fair comparison.

We implement \ToolName{} using LangChain~\cite{chase_langchain_2022} and vLLM~\cite{kwon2023efficient}. To reduce hallucinations and guarantee syntactic correctness, we adopt a two-stage generation pipeline. Instead of emitting annotated code directly, the model first generates a structured insertion plan, which we then apply deterministically to the source code.

To encourage diversity among candidates, we use sampling with temperature $=1.0$ and top-$p=1.0$. To ensure these plans are reliably parsable, we enforce a strict JSON schema using the structured output mechanisms provided by the underlying frameworks. For the RAG component, we configure the retriever to return the top $k=4$ relevant documents as context.

\section{Case Study}
\label{app:case}

\begin{figure}[t]
    \centering
    \includegraphics[width=\linewidth]{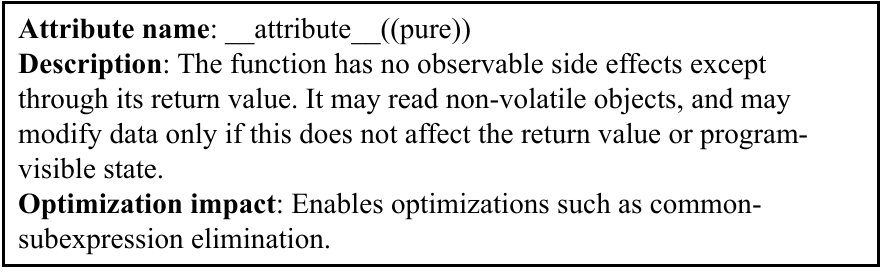}
    \caption{\textbf{Case \textrm{I}.} \ToolName{} retrieves the usage pattern for the \texttt{pure} attribute and inserts it, enabling the compiler to optimize the call.}
    \label{fig:case1b}
    %\vspace{-3mm}
\end{figure}

As discussed in Section~\ref{methodology}, \ToolName{} significantly mitigates hallucinations in compiler hint generation by grounding decisions in retrieved documentation. We illustrate this process with several representative examples.

% \begin{figure*}[t]
%     \centering
%     \small
%     \begin{subfigure}{\linewidth}
%         \includegraphics[width=\linewidth]{paper-llm4opt/Figures/case1.pdf}
%         \caption{Optimized Code} 
%         \label{fig:case1a}
%     \end{subfigure}
%     \hfill 
%     \begin{subfigure}{\linewidth}
%         \includegraphics[width=\linewidth]{paper-llm4opt/Figures/case1_tr.pdf}
%         \caption{Retrieved Knowledge} 
%         \label{fig:case1b}
%     \end{subfigure}
%     \caption{\textbf{Case \textrm{I}.} (a) The original code is dominated by a deeply-nested loop nest in the dynamic programming kernel. (b) \ToolName{} retrieves the usage pattern for the \texttt{pure} attribute and inserts it, enabling the compiler to optimize the call.}
%     \label{fig:case1}
% \end{figure*}

% \begin{figure*}[t]
%     \centering
%     \small
%     \includegraphics[width=\linewidth]{paper-llm4opt/Figures/case1.pdf}
%     \caption{\textbf{Case \textrm{I}.} The original code is dominated by a deeply-nested loop nest in the dynamic programming kernel.}
%     \label{fig:case1a}
% \end{figure*}

%\jiang{an example}
\subsection{Case \textrm{I}}

\begin{figure}[t]
    \centering
   % \small
    \includegraphics[width=0.95\textwidth]{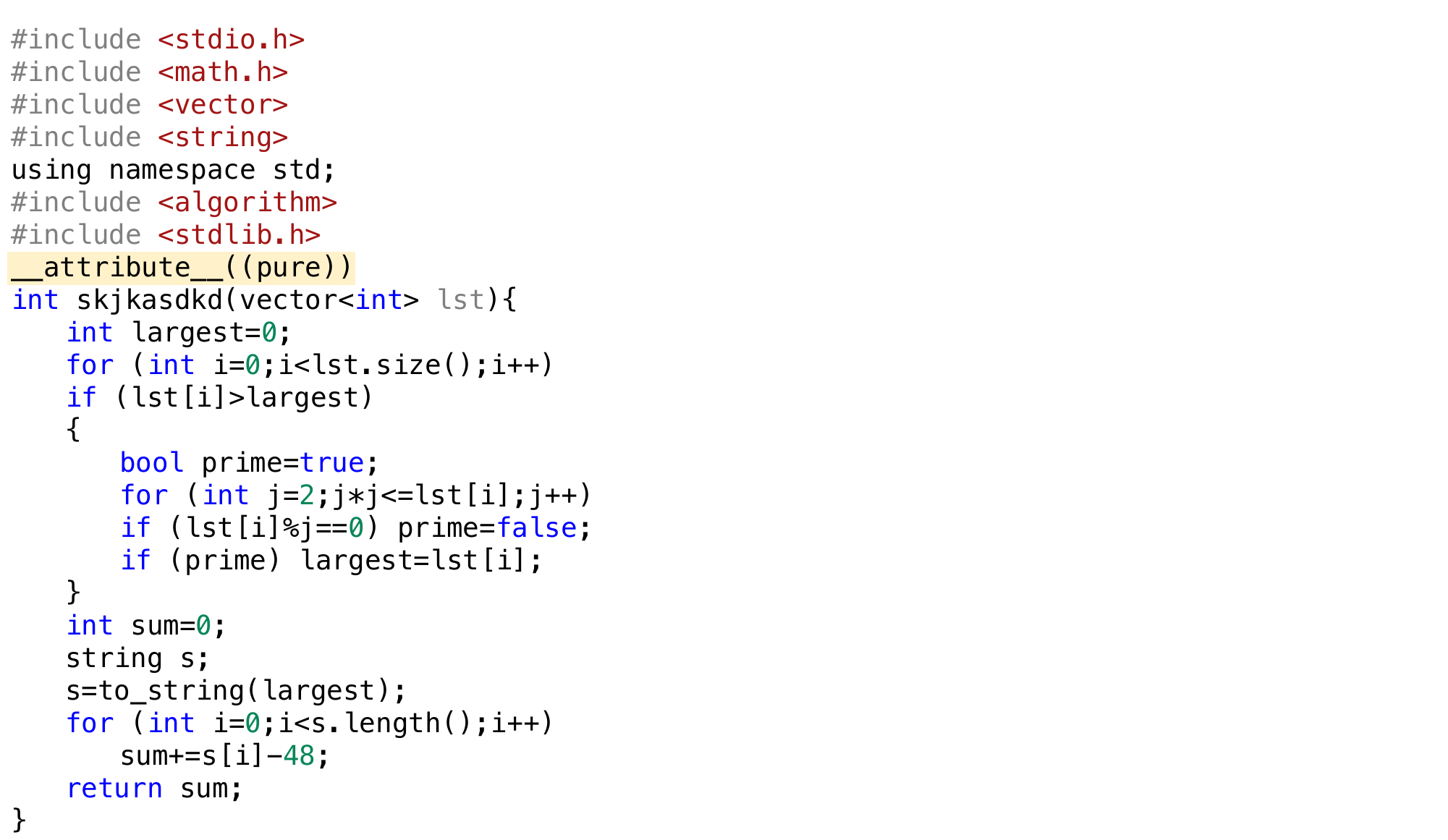}
    \caption{\textbf{Case \textrm{I}.} The original code is dominated by a deeply-nested loop nest in the dynamic programming kernel.}
    \label{fig:case1a}
\end{figure}

Consider the source program in Figure~\ref{fig:case1a}, which contains a computationally intensive function called within a loop. \ToolName{} first parses the code structure and queries the knowledge base. Figure~\ref{fig:case1b} displays one retrieved result: a canonical usage example of the \texttt{pure} attribute, explicitly stating that it applies to functions with no side effects.

Grounded by this context, the LLM correctly infers that the target function depends solely on its arguments and modifies no global state. It then generates a plan to insert \texttt{\_\_attribute\_\_((pure))}. This annotation explicitly informs the compiler that the function is side-effect-free, enabling aggressive optimizations such as loop-invariant code motion (hoisting the function call out of the loop) and common subexpression elimination, thereby significantly reducing runtime overhead. Remarkably, this single attribute injection results in a 325$\times$ speedup, primarily by eliminating redundant computations inside the hot loop.

% \begin{figure*}[t]
%     \centering
%     \begin{subfigure}{\linewidth}
%         \includegraphics[width=\linewidth]{paper-llm4opt/Figures/case2.pdf}
%         \caption{Optimized Code} 
%         \label{fig:case2a}
%     \end{subfigure}
%     \hfill 
%     \begin{subfigure}{\linewidth}
%         \includegraphics[width=0.48\linewidth]{paper-llm4opt/Figures/case2_tr1.pdf}
%         \includegraphics[width=0.48\linewidth]{paper-llm4opt/Figures/case2_tr2.pdf}
%         \caption{Retrieved Knowledge} 
%         \label{fig:case2b}
%     \end{subfigure}
%     \caption{\textbf{Case \textrm{II}.} (a) The original code contains a redundant function call inside a loop. (b) \ToolName{} retrieves function-level attributes and loop-level OpenMP directives, enabling parallel execution and vectorization of the core computation.}
%     \label{fig:case2}
% \end{figure*}

% \begin{figure*}[h]
%     \centering
%     \small
%     \includegraphics[width=\linewidth]{paper-llm4opt/Figures/case2.pdf}
%     \caption{\textbf{Case \textrm{II}.} The original code contains a redundant function call inside a loop.}
%     \label{fig:case2a}
% \end{figure*}

\begin{figure*}[t]
    \centering
   % \small
    \includegraphics[width=0.95\linewidth]{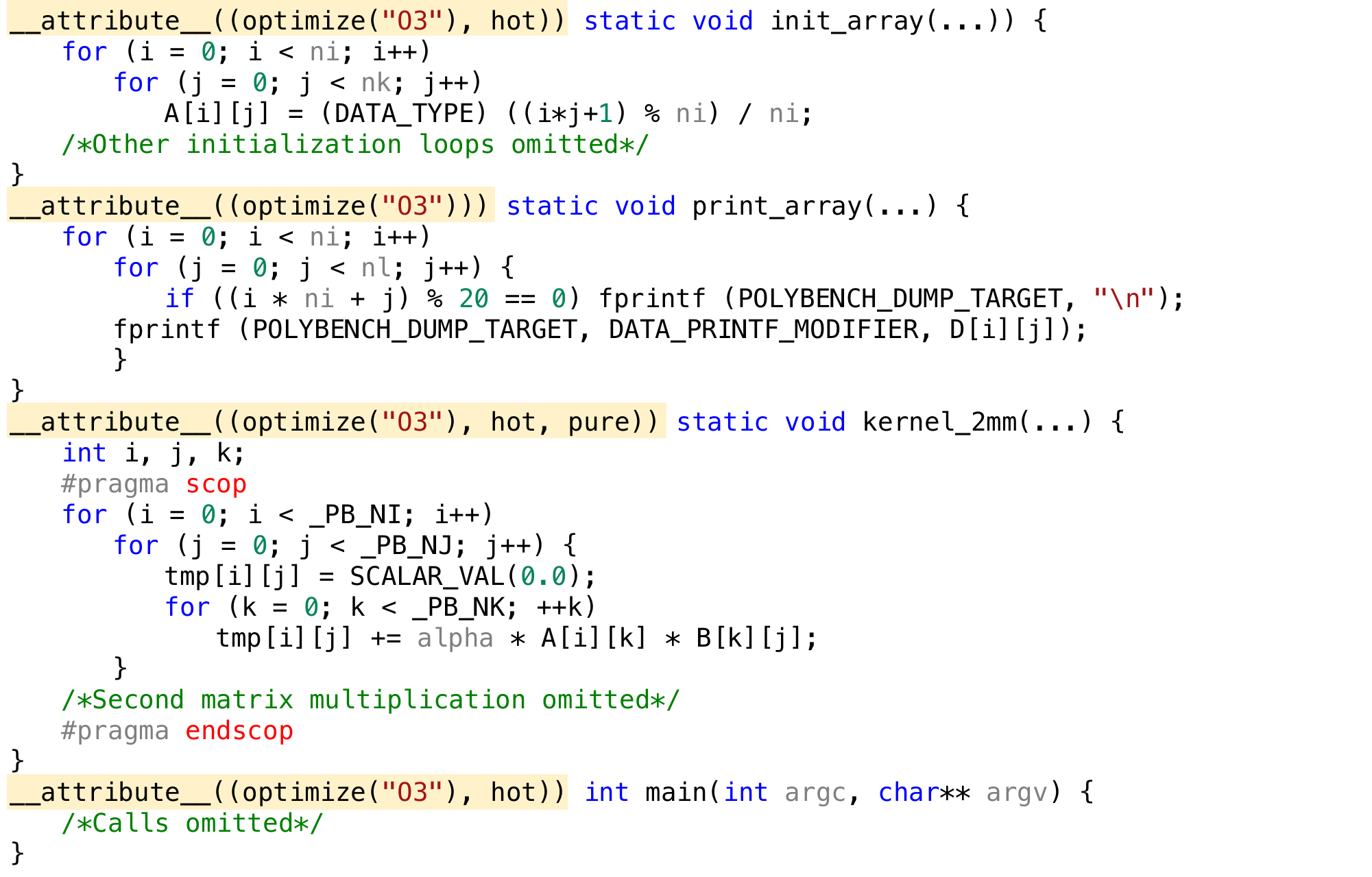}
    \caption{\textbf{Case \textrm{II}.} The original code contains a redundant function call inside a loop.}
    \label{fig:case2a}
\end{figure*}

\subsection{Case \textrm{II}}
\begin{figure}[t]
    \centering
    \includegraphics[width=\linewidth]{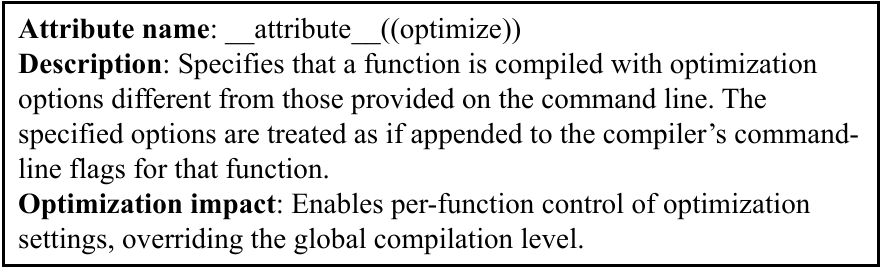}

  % \vspace{2mm}

    \includegraphics[width=\linewidth]{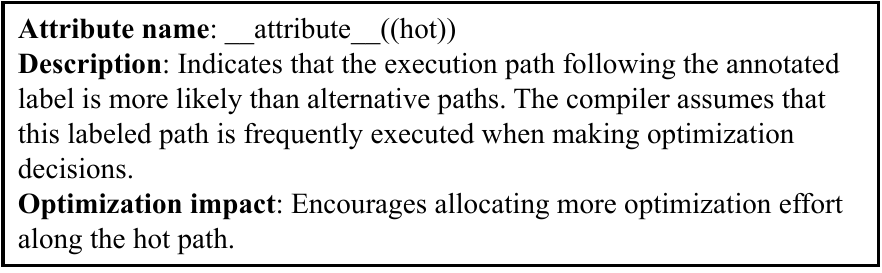}

    \caption{\textbf{Case \textrm{II}.} \ToolName{} retrieves compiler optimization attributes for compute-intensive kernels and applies them to prioritize optimization of the main execution path.}
    \label{fig:case2b}
   % \vspace{-3mm}
\end{figure}

Consider the source program in Figure~\ref{fig:case2a}, whose computation is dominated by two consecutive matrix multiplications in the \texttt{2mm} kernel. \ToolName{} first parses the program and then queries the knowledge base for relevant compiler attributes. Figure~\ref{fig:case2b} displays retrieved knowledge including \texttt{optimize("O3")} and \texttt{hot}, which convey optimization priority in compute-intensive regions.

Grounded by this context, the LLM synthesizes a plan that assigns \texttt{\_\_attribute\_\_((optimize("O3"),~hot))} to the kernel as well as other routines on the main execution path. These annotations inform the compiler that the functions lie on the performance-critical execution path, enabling aggressive optimizations and prioritizing code generation. The kernel is additionally marked as \texttt{pure}, which reduces side-effect-related constraints and allows more effective instruction scheduling and register allocation. The attributed version achieves an 89$\times$ speedup by reducing computation and memory-access overhead in the core kernel.

% \begin{figure*}[t]
%     \centering
%     \small
%     \begin{subfigure}{\linewidth}
%         \includegraphics[width=\linewidth]{paper-llm4opt/Figures/case3.pdf}
%         \caption{Optimized Code} 
%         \label{fig:case3a}
%     \end{subfigure}
%     \hfill 
%     \begin{subfigure}{\linewidth}
%         \includegraphics[width=\linewidth]{paper-llm4opt/Figures/case3_tr.pdf}
%         \caption{Retrieved Knowledge} 
%         \label{fig:case3b}
%     \end{subfigure}
%     \caption{\textbf{Case \textrm{III}.} (a) The original code consists of a compute-intensive initialization routine and the LU factorization kernel. (b) \ToolName{} retrieves \texttt{hot} and \texttt{cold} attributes but incorrectly marks a compute-intensive routine as \texttt{cold}.}
%     \label{fig:case3}
% \end{figure*}

% \begin{figure*}[h]
%     \centering
%     \small
%     \includegraphics[width=\linewidth]{paper-llm4opt/Figures/case3.pdf}
%     \caption{\textbf{Case \textrm{III}.}The original code consists of a compute-intensive initialization routine and the LU factorization kernel.}
%     \label{fig:case3a}
% \end{figure*}

\subsection{Case \textrm{III}}
\label{case3}

\begin{figure*}[t]
    \centering
    \small
    \includegraphics[width=0.95\linewidth]{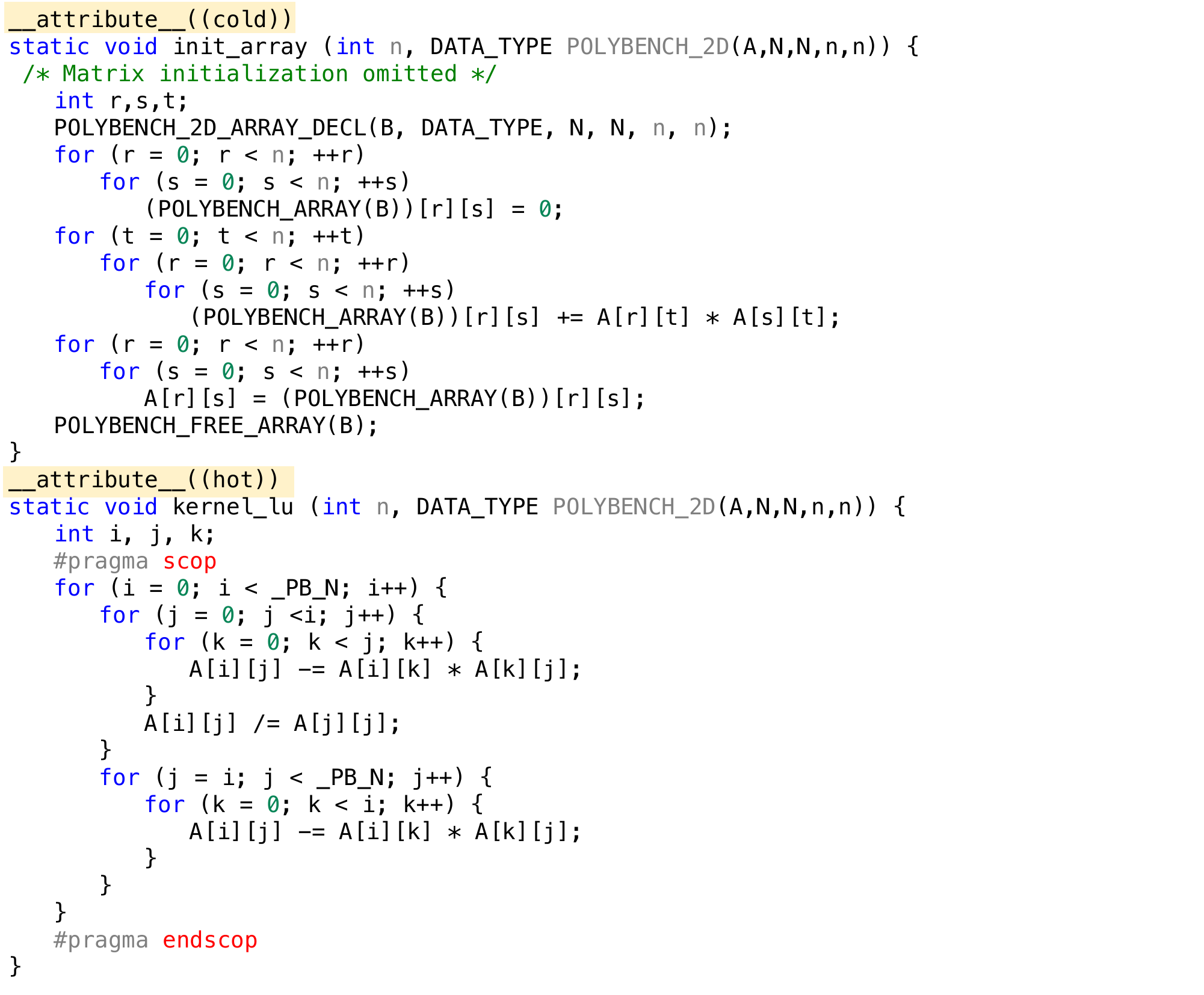}
    \caption{\textbf{Case \textrm{III}.}The original program includes a compute-intensive initialization phase followed by an LU factorization kernel}
    \label{fig:case3a}
\end{figure*}

\begin{figure}[t]
    \centering
    \includegraphics[width=\linewidth]{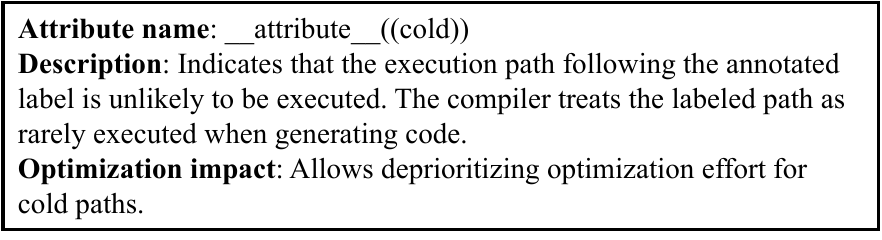}
    \caption{\textbf{Case \textrm{III}.} \ToolName{} retrieves the \texttt{cold} attribute and applies it to the initialization routine, deprioritizing optimization in a compute-intensive region.}
    \label{fig:case3b}
  %  \vspace{-3mm}
\end{figure}

Consider the program in Figure~\ref{fig:case3a}, which consists of a compute-intensive initialization routine followed by the factorization kernel. \ToolName{} profiles the execution to identify performance-critical regions, determines that the LU kernel is the main bottleneck, and queries the knowledge base. Figure~\ref{fig:case3b} displays the retrieved \texttt{cold} attribute, indicating expected execution frequency and allowing the compiler to adjust optimization effort accordingly.

Grounded in this context, the LLM marks \texttt{kernel\_lu} as \texttt{\_\_attribute\_\_((hot))} and assigns \texttt{\_\_attribute\_\_((cold))} to the initialization routine. However, in this case, the initialization phase performs substantial computation and accounts for a significant portion of the total runtime, even though it is executed only once. The \texttt{cold} attribute, therefore, biases the compiler toward more conservative code generation in a compute-intensive region, leading to increased execution time, while the \texttt{hot} attribute on the LU kernel provides little benefit due to inherent loop dependences. 

Moreover, this case reveals instability in the feedback loop. Since performance is evaluated only at the program level, the feedback is dominated by coarse-grained runtime noise, making it impossible to attribute the slowdown to the initialization phase and preventing effective refinement in subsequent optimization attempts.
As a result, it fails to complete within the time limit because it incorrectly marks the performance-critical routine as non-critical.

\section{Addiditional Experiments}
\subsection{Inference Time Scaling}
To further explore the potential of direct prompting, we investigated the effect of test-time compute scaling with CoT+Few-shot prompt by varying the output token budget. We observed that reducing the budget significantly degrades optimization performance, confirming that reasoning space is essential. However, increasing the budget beyond our default setting(8k) yielded diminishing returns, indicating that the LLM's inherent reasoning capacity for this task saturates at this level. To the best of our knowledge, our current setup represents the strongest possible baseline for pure prompting. The fact that HINTPILOT still outperforms this saturated baseline underscores our contribution to code optimization.

\begin{table}[htbp]
\centering
\begin{tabular}{lcc}
\hline
Token Budget & HumanEval & Polybench \\
\hline
4k  & 1.28$\times$ & 1.18$\times$ \\
16k & 1.65$\times$ & 1.16$\times$ \\
8k (HINTPILOT) & 3.53$\times$ & 2.10$\times$ \\
\hline
\end{tabular}
\caption{Performance comparison under different token budgets}
\end{table}

\subsection{Time Overhead}
We have added an additional timing study to quantify time overhead: we report the average wall-clock time per round and the end-to-end runtime under our default setting, with a breakdown into (i) hints generation, (ii) execution-guided self-refinement. We currently test Qwen3-Coder-Plus and Codestral-22B-v0.1 with default settings in the following Tables.

\begin{table}[t]
\centering
\scriptsize
\tiny
\begin{tabular}{l l c c}
\hline
Dataset & Model & Hint Gen. (s) & Exec.-Guided (s) \\
\hline
PolyBench & Qwen3-Coder-Plus        & 26.91 & 40.12 \\
PolyBench & Codestral-22B-v0.1      & 10.23 & 60.33 \\
HumanEval\_CPP & Qwen3-Coder-Plus   & 18.61 & 6.06 \\
HumanEval\_CPP & Codestral-22B-v0.1 & 7.17  & 5.63 \\
\hline
\end{tabular}
\caption{Runtime (seconds) of hint generation and execution-guided self-refinement.}
\label{tab:runtime}
\end{table}

\subsection{Impact of $T$ and $N$}
Our initial choice of $T$ and $N$ was a balance between optimization performance and LLM API budgetary considerations. We have also conducted an extensive scaling study. These results support our hypothesis that iterative refinement is beneficial, but it must be paired with sufficiently diverse candidate generation (larger $N$) to avoid local optima. And increasing $T$
slightly decreases performance when $N$ is small, possibly due to noise accumulation in the feedback signals.
\begin{figure}[h]
    \centering
    \includegraphics[width=\linewidth]{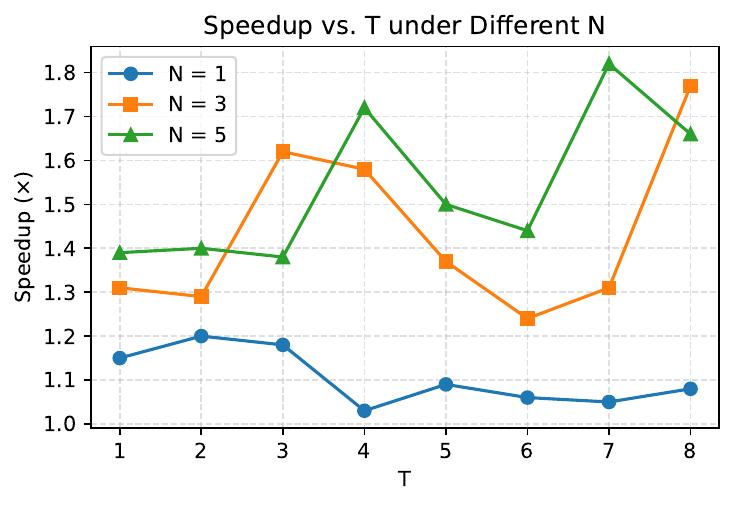}
    \caption{Speedup performance scaled with $T$ and $N$.}
    \label{fig:scale_T_N}
\end{figure}

\subsection{Real-world Application}
we further extended experiments to three industry-standard C++ repositories: Redis, LevelDB, and spdlog. These widely used systems span multiple domains, including in-memory databases, storage engines, and logging systems, and each provides a comprehensive performance test suite for quantitative assessment. For each project, we first used perf to profile hot functions and files, and then applied HINTPILOT. In the camera-ready version, we will further broaden the evaluation to additional real-world projects to strengthen coverage. Here, Perf Improvement reports the maximum and average improvement ratios across all test cases:

\begin{table}[t]
\centering
\scriptsize
\begin{tabular}{lcc}
\hline
Repo & Perf. Improvement $\uparrow$ (Max) & Perf. Improvement $\uparrow$ (Avg) \\
\hline
LevelDB & 54.00\% & 9.20\% \\
spdlog  & 48.00\% & 11.59\% \\
Redis   & 10.42\% & 0.62\% \\
\hline
\end{tabular}
\caption{Maximum and average performance improvements across repositories.}
\label{tab:perf_improvement}
\end{table}

\ToolName{} achieves up to 54\% improvement on LevelDB and 48\% on spdlog, with over 22\% of LevelDB test cases exceeding a 10\% speedup. While the average gain on Redis is modest, we still observe a maximum speedup of 8.10\%. These results indicate that, even with fewer degrees of freedom than full code rewriting, \ToolName{} can meaningfully address optimization demands in large-scale codebases.

Crucially, because \ToolName{} operates through non-invasive annotations, it is well-suited to real-world engineering settings where developers are often reluctant to undertake extensive refactoring due to restricted access, high verification costs, and the risk of regressions in legacy systems.

\subsection{Cross compiler performance}
While our primary evaluation focused on GCC due to its industrial prevalence, we have conducted an additional pilot study using Clang on HumanEval with O3. The results show \ToolName{} achieved an average speedup of 1.56x over O3 and 1.40x over Ofast with Clang compiler, demonstrating the compiler-agnostic nature of our architecture. We will add experiments on other compilers in the camera-ready version.

\section{Prompt Templates}
\label{app:prompts}

This appendix reports the exact prompt templates used for Zero-shot,
Chain-of-Thought (CoT), and CoT + few-shot settings.
All configurations share identical task instructions and strict
JSON-only output constraints, differing only in whether private
deliberation (CoT) and in-context examples (few-shot) are provided.

\subsection{System Prompt}
\begin{lstlisting}[breaklines=true]
You are a compiler attribute advisor.
Your goal: recommend only semantics-preserving GCC/Clang attributes
that can potentially accelerate program execution time.
\end{lstlisting}
\subsection{Task Instruction and Constraints}
\begin{lstlisting}[breaklines=true]
Output requirements:
- Strictly return a single valid JSON object (UTF-8),
  with no Markdown, no code fences, and no extra text.
- Do not include comments or unused / extra keys.
- Return ONLY valid JSON.

Constraints:
- Recommend only semantics-preserving edits. If safety is uncertain,
  lower confidence or skip.
- Use mainstream GCC/Clang attributes, e.g.:
  * function: hot, cold, flatten, noinline, always_inline, malloc,
    pure, const (when safe)
  * variable: aligned(...), visibility(...)

- Loops:
  * OpenMP only if no loop-carried dependencies
  * Use collapse(N) only for perfectly nested independent loops
  * Reductions only when clearly safe

- Insert attributes before variables/functions.
- Multiple hints/candidates allowed.
- JSON output only; no hidden reasoning.
\end{lstlisting}

\subsection{Zero-shot Prompt}
\begin{lstlisting}[breaklines=true]
<task instruction>

Return ONLY a JSON object:
{"code": "<the FULL transformed source code>"}
\end{lstlisting}

\subsection{Chain-of-Thought (CoT) Prompt}
\begin{lstlisting}[breaklines=true,
    breakatwhitespace=false,
    columns=fullflexible,
    keepspaces=true]
Deliberate privately:
- Reason step by step about safety, dependencies, aliasing,
  reductions, side effects, and OpenMP semantics.

<task instruction>

CODE WITH MARKERS:
{parse_json}

Output ONLY JSON:
{"hints":[{"symbol":"<name>","kind":"function|global",
"line":<int>,"col":<int>,"reason":"<str>",
"candidates":[{"attr":"__attribute__((...))|#pragma",
"reason":"<str>"}]}]}
\end{lstlisting}

\subsection{CoT + Few-shot Prompt}
\begin{lstlisting}[breaklines=true,
    breakatwhitespace=false,
    columns=fullflexible,
    keepspaces=true]
You are a compiler attribute advisor.
Your goal: recommend only semantics-preserving GCC/Clang attributes.

Deliberate privately about safety and semantic preservation.

Parsed attribute positions (JSON):
{parse_json}

Examples omitted for page limit.

<task instruction>

Output format:
{"hints":[{"symbol":"<name>","kind":"function|global", 

"line":<int>,"col":<int>,"reason":"<str>",
"candidates":[{"attr":"__attribute__((...))|#pragma", 

"reason":"<str>"}]}]}
\end{lstlisting}